\documentclass[twoside,11pt]{article}
\usepackage{jair}
\usepackage{url}
\usepackage{latexsym}
\usepackage{amsmath}
\usepackage{multirow}
\usepackage{graphicx}
\usepackage{amsfonts}
\usepackage{bm}
\usepackage{amsmath}
\usepackage{amssymb}
\usepackage{color}
\usepackage{version}
\usepackage[square,sort,comma,numbers]{natbib}
\usepackage{hyperref}
\hypersetup{%
setpagesize=false,
bookmarksnumbered=true,%
bookmarksopen=true,%
colorlinks=true,%
linkcolor=blue,
citecolor=red,
}
\usepackage{amsthm}
\newtheorem{assumption}[]{Assumption}

\ShortHeadings{Personalized Visited-POI Assignment to Individual Raw GPS Trajectories}
{Suzuki, Suhara, Toda \& Nishida}
\firstpageno{1}

\title{Personalized Visited-POI Assignment \\to Individual Raw GPS Trajectories}
\author{\name Jun Suzuki\thanks{Current affiliation: Tohoku University}\\
      \addr  NTT Communication Science Laboratories, NTT Corporation\\
      \name Yoshihiko Suhara\thanks{Current affiliation: Megagon Labs} \email\\
      \name Hiroyuki Toda \email\\
      \addr  NTT Service Evolution Laboratories, NTT Corporation\\
      \name Kyosuke Nishida \email\\
      \addr NTT Media Intelligence Laboratories, NTT Corporation\\
      }
\begin{document}
\maketitle

\begin{abstract}
Knowledge discovery from GPS trajectory data is an important topic in several scientific areas, including data mining, human behavior analysis, and user modeling.
This paper proposes a task that assigns personalized visited-POIs.
Its goal is to estimate fine-grained and pre-defined locations (i.e., points of interest (POI)) that are actually visited by users and assign visited-location information to the corresponding span of their (personal) GPS trajectories.
We also introduce a novel algorithm to solve this assignment task.
First, we exhaustively extract stay-points as candidates for significant locations using a variant of a conventional stay-point extraction method.
Then we select significant locations and simultaneously assign visited-POIs to them by considering various aspects, which we formulate in integer linear programming.
Experimental results conducted on an actual user dataset show that our method achieves higher accuracy in the visited-POI assignment task than the various cascaded procedures of conventional methods.
\end{abstract}

\section{Introduction}
\label{sec:introduction}
The availability of personal spatial-temporal data continues to rapidly increase.
This is because personally equipped mobile devices such as smartphones have become ubiquitous and are generally equipped with GPS devices that can constantly record the trajectories of latitude-longitude positions.
This situation provides opportunities to discover valuable knowledge from such personal spatial-temporal data.
In fact, knowledge discovery tasks from GPS trajectory data have already been proposed, such as predicting movement destinations \cite{Ashbrook:2003,Ying:2011}, recommending points of interest (POIs) around locations \cite{Lian:2011,Lian:2014,Shaw:2013,Ye:2011,Yuan:2013}, and recognizing individual mobility \cite{Zheng:2010,Zheng:2011a}.
Unfortunately, mining knowledge from raw GPS trajectory data is not a straightforward task
because such data are merely a series of real-valued positions and timestamps.
We need to link the GPS trajectory data to the actual world's semantic information, including that for locations, events, transportation modes, user activities, and personal preferences.
Hence, GPS trajectory mining is attracting a great deal of attention in the geospatial data mining community.

Following this line of study, we discuss a similar but new GPS trajectory data mining challenge called a personalized visited-POI assignment task.
Its goal is to estimate the fine-grained locations that a user actually visits and assign the user's (personal) GPS trajectories to them.
A visited-POI indicates not only location history but also user habits and preferences.
At least two crucial applications can be inferred from personalized visited-POI assignment tasks.
One is the automatic construction of a user's personal lifelogs from his/her personal spatial-temporal data based on GPS trajectories.
Since annotating all GPS trajectory data by hand is unmanageable because the amount of GPS trajectory data is excessive,
a personalized visited-POI assignment system might significantly reduce the annotation cost of constructing personal spatial-temporal lifelogs.
Another application is  POI recommendation systems \cite{Shaw:2013,Ye:2011,Yuan:2013}, which basically
use personal POI histories and general POI patterns that can be essentially estimated by our visited-POI assignment method.

The automatic assignment of personalized visited-POIs to individual raw GPS trajectories is challenging.
For example, several researchers have tackled the task of understanding user-specific activity from raw GPS trajectory data \cite{Krumm:2007,Li:2008b,Zheng:2011a}.
In their work, they assumed that a {\it stay-point}, which is a spot where the user remains over a constant time-span within a certain area, is a unit that pertains to a meaningful location of a user.
Since they assume that stay-points are accurately extracted, detecting significant locations from them remains an open question.
This is because the stay times of significant locations vary, and a stay-point extraction algorithm with just one fixed parameter cannot extract all of the significant locations without errors.

We propose a novel framework to solve this task.
We first enumerate all of the possible candidates of the time-spans for a visited-POI using an approach that resembles the conventional stay-point extraction method with a very conservative parameter setting.
Our approach extracts the true time-spans at which the user actually stayed in the visited-POIs from a GPS trajectory with very high recall.
Thus, we can select the significant time-spans from the extracted stay-points and simultaneously assign their visited-POIs.
At that time, we need to capture several different aspects of the personal and general human behavior for accurate estimation.
For example, we must consider the validity of the extracted stay-points, the likelihood of the visited-POI assignments, the validity of the visited-POI relations, and the validity of the visited-POI sequence length.
To simultaneously take these heterogeneous variables into account, we formulate the challenge as an instance of a combinatorial optimization problem and solve it as integer linear programming (ILP).

Note that one aspect about which we must be cautious is the task settings that are related to how we obtain and incorporate personal preferences and information.
People are often unwilling to upload their GPS trajectories (justifiably so) and their location (visited-POIs in our case) histories to servers because of privacy concerns \cite{Xue:2013}.
Thus, we assume that users' personal information, including their visited-POI histories, cannot be aggregated in one system of a service.
Instead, we assume that individual GPS histories are collected in a user's own storage.
This means that the personal information of other users cannot be incorporated for building a visited-POI assignment system.
We restrict our reach to a user's personal information for estimating his/her visited-POI assignments.
We believe this is a very realistic task setting.

The following are the contributions of this paper:
\begin{enumerate}
\item We revisit and evaluate the conventional stay-point extraction algorithm and provide an appropriate strategy adjusted to the personalized visited-POI assignment task.
\item We propose a general framework that efficiently and accurately detects significant time-spans that can be assigned to visited-POIs.
\end{enumerate}

The rest of the paper is organized as follows:
We first describe the definitions and the problem setting of our personalized visited-POI assignment task in Section \ref{sec:problem_setting} and
discuss related work in Section \ref{sec:related_work}.
We present the proposed method to tackle the personalized visited-POI assignment task in Section \ref{sec:proposed_method} and
describe our experiments and results in Section \ref{sec:evaluation}.

\section{Definition of Personalized Visited-POI Assignment}
\label{sec:problem_setting}

\begin{figure}[t]
\centering
\includegraphics[width=14cm, clip]{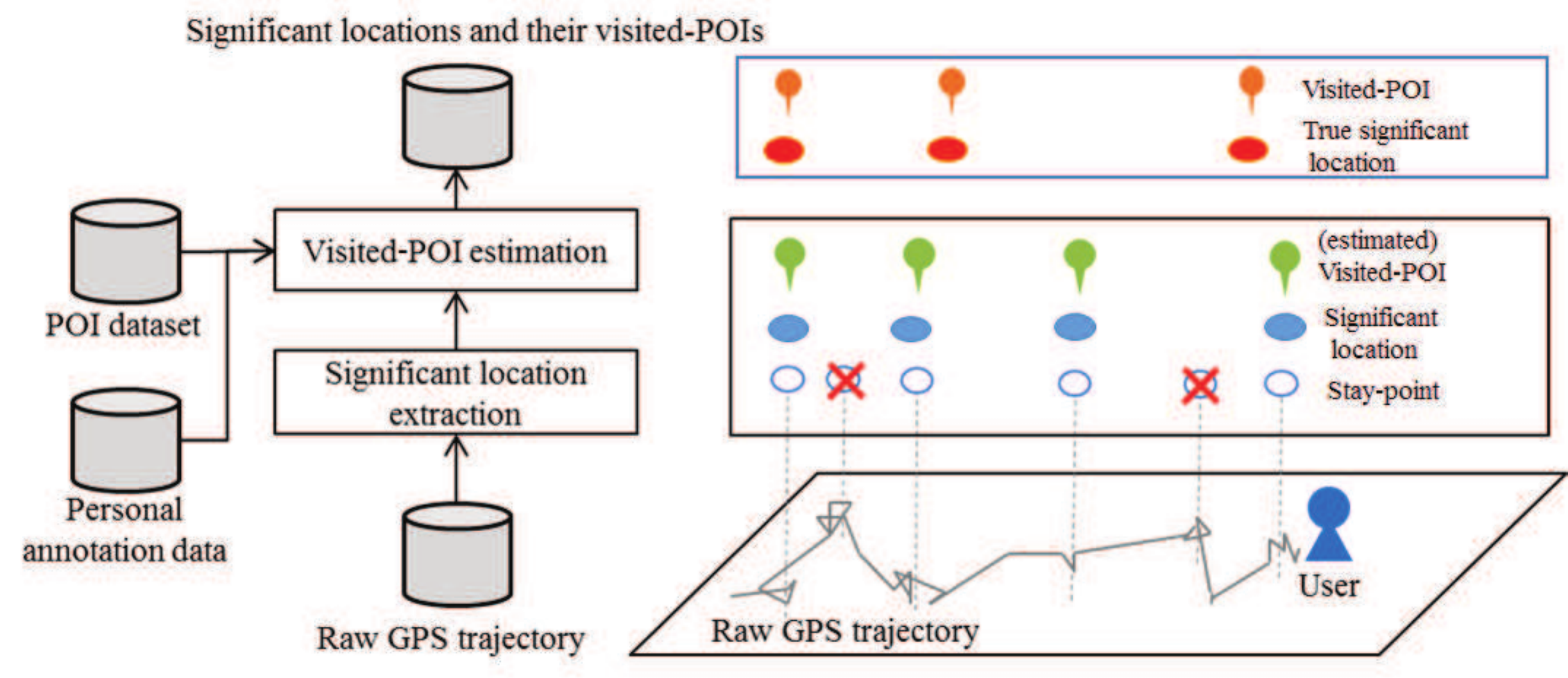}
\caption{Problem setting in this paper}
\label{fig:system_overview}
\end{figure}

This section describes an overview and the definition of our proposed personalized visited-POI assignment task.

\subsection{Preliminary: Terms and Notations}
For clarity, we first introduce some technical terms and provide their definitions used in this paper.

\textbf{Track-point}:
We refer to a single data point obtained from a positioning system such as GPS as a track-point
and assume that a track-point consists of a triplet: the longitude, the latitude, and the actual timestamp to be recorded.
Formally, let $lng$, $lat$ and $ts$ respectively be the longitude, the latitude, and the actual timestamp.
Then, a track-point $G$ is written as $G=(lng, lat, ts)$.

\textbf{Session}:
A session is a sequence of continuous track-points.
Let $G_{t}$ and $G_{t+1}$ represent successive track-points at times $t$ and $t+1$, respectively.
Then session $S$ is defined as $S=(G_1, G_2, \dots, G_T)$, where $T$ represents the number of track-points in the session.
Our study obtains one session from a sequence of GPS trajectory for a day, where
a session always starts at 0:00 midnight and ends at 23:59:59 to simplify the task definition.

\textbf{Stay-point}:
A stay-point~\cite{Kang:2004,Zheng:2009} is a sub-sequence of the session $S$ at which the user remains over a certain pre-defined period in the same area.
Formally, a stay-point $sp$ consists of successive track-points in a session $S$, namely, $sp=(G_i, G_{i+1}, \dots, G_j)$, where $i$ and $j$ are the indices of the starting and ending points of the stay-point, that is, $1 \leq i \leq j \leq T$.
Hereafter, the relation between a sub-sequence $A$ in a sequence $B$ is denoted as $A \subseteq B$.
Thus, the relation between a stay-point $sp$ in the session $S$ can be written as $sp \subseteq S$.

In this paper, each stay-point has five additional attributes, namely, the center location ($lng$, $lat$) of the sequence of the track-points in the stay-point, start timestamp ${bt}$, end timestamp ${et}$, and stay-time ${st}$, all of which are easily calculated from the sequence of track-points in the stay-point.
Suppose $sp_k = (G_i, G_{i+1}, \dots, G_j)$ represents the $k$-th stay-point in a session $S$.
We refer to the center locations of the longitude, the center location of the latitude, the start timestamp, the timestamp, and the stay-time of $sp_k$ as $sp_k.{lng}$, $sp_k.{lat}$, $sp_k.{bt}$, $sp_k.{et}$, and $sp_k.{st}$, respectively;
\begin{equation}
sp_k.{lng} = \frac{1}{j-i+1}\sum^{j}_{t=i} G_{t}.lng
 ,
\end{equation}
\begin{equation}
sp_k.{lat} = \frac{1}{j-i+1}\sum^{j}_{t=i} G_{t}.lat
 ,
\end{equation}
\begin{equation}
sp_k.{bt}  = G_i.{ts},
\end{equation}
\begin{equation}
sp_k.{et}  = G_j.{ts}
 ,
\end{equation}
\begin{equation}
sp_k.{st}  = G_j.{ts} - G_i.{ts}
 .
\end{equation}

\textbf{Point of Interest (POI)}\footnote{See \url{https://en.wikipedia.org/wiki/Point\_of\_interest} for a quick reference.}:
A POI~\cite{Lian:2011,Shaw:2013,Ye:2011,Yuan:2012a,Yuan:2013} is a place in which people may be interested.
Suppose we have a pre-defined set of POIs (POI database) denoted as $\mathcal{P}$.
This paper defines that each POI consists of the following four attributes; its own name, its category, and its location (longitude, latitude) information.
Let $poi_k$ represent the $k$-th POI in $\mathcal{P}$, that is, $poi_k\in\mathcal{P}$.
We refer to the name, category, longitude and latitude of $poi_k$ as $poi_k.name$, $poi_k.cat$, $poi_k.lng$ and $poi_k.lat$, respectively.

 Then we define two technical terms, {\it visited-POI} and {\it significant location}, that are the keys for describing our proposed task.

\textbf{Visited-POI}:
 A visited-POI is a POI that the user has actually visited in the real world.
 Note that a visited-POI also consists of identical attributes as POI, namely,
 the name, category, longitude and latitude.

\textbf{Significant location}:
Similar to the stay-point, a significant location is a sub-sequence of session $S$ while the user actually visited to a POI.
In other words, this is a sequence of track-points in $S$ that can be assigned to a corresponding visited-POI.
We define that significant location also consists of identical attributes as a stay-point.
This means that each significant location also has five additional attributes; the center location of the track-points, the start timestamp, the end timestamp, and the stay-time of the significant location as well as stay-point.

\subsection{Task Definition and Assumptions}
In this section, we propose a task called {\it personalized visited-POI assignment task}.
With the terms and definitions explained in the previous section, this section describes our task definition.
Fig.~\ref{fig:system_overview} shows an overview of our task.

We first explain a few required assumptions for setting our task a meaningful study.
The first assumption addresses about the relation between stay-point and significant location.
\begin{assumption}
We assume that each significant location is always a stay-point.
\end{assumption}
This assumption means that we infer that users always stay for a certain duration of time if they visit a POI.
In other words, the task proposed in this paper only considers the POIs at which the users stay for a particular  period of time, i.e., for one minute.
This assumption is not unrealistic since users generally have their own purposes to do for {\it visiting} a POI.
Note that the above assumption does not mention that a stay-point is always a significant location.
 For example, the situations of being stuck in traffic, waiting at railroad crossings, and talking on cell phones in open spaces without changing locations are typical and intuitive examples of stay-points that are not significant locations.
In addition, Fig.~\ref{fig:system_overview} also illustrates the relation among stay-points, significant locations, and visited-POIs, as defined in the previous section.

The second assumption is about the pre-defined POI database used in our paper.
\begin{assumption}
We assume that a (common) POI database $\mathcal{P}_{c}$ is obtained from a conventional location-based service (e.g., Foursquare\footnote{\url{https://foursquare.com/}}).
We also assume that each user may have his/her own personal POI database $\mathcal{P}_{p}$, which includes his/her home, office, and places in which he/she is specifically interested.
\end{assumption}
This assumption indicates that all users might have their own POI databases $\mathcal{P}_{p+c} = \mathcal{P}_{p} \cup \mathcal{P}_{c}$.
Note that if user does not have his/her own POI database $\mathcal{P}_{p}$, then $\mathcal{P}_{p+c} = \mathcal{P}_{c}$.

The following is an assumption about how we determine the significant locations and visited-POIs.
\begin{assumption}
In this paper, we assume that the manually annotated visited-POIs and the significant locations by users themselves are true visited-POIs and significant locations.
\end{assumption}
This assumption is derived because precisely identifying the places actually visited by users is difficult.
Therefore, we trust the annotations of users.
Moreover, since the granularity of POI annotation deeply depends on users, significant locations also deeply depend on users intentions about where they believe they visited.

The following is the assumption about the relation among significant locations.
\begin{assumption}
The duration of visits, in other words, the duration of significant locations, never overlaps.
\end{assumption}
This assumption is reasonable and realistic since a user cannot physically visit more than two places at one time.
This assumption reflects our annotation scheme; users cannot annotate overlapping visited-POIs and significant locations since they are not expected to simultaneously visit more than two POIs.
Given two stay-points $sp_m$ and $sp_n$ in a session $S$, namely, $sp_m \subseteq S$ and $sp_n \subseteq S$,
$sp_m$ and $sp_n$ are disjoint (non-overlap) stay-points in terms of the duration of the stays if the relation $G_{a} \neq G_{b}$ holds for all $G_{a} \in sp_m$ and $G_{b} \in sp_n$.
We represent the relation of such non-overlapping two sub-sequences as $sp_n \cap sp_m = \emptyset$.

\textbf{Personalized visited-POI assignment task}:
Let $(sp_n,poi_n)$ represent the $n$-th significant location with corresponding visited-POI, where $sp_n$ and $poi_n$ represent the $n$-th significant location and its corresponding visited-POI, respectively.
Then given a (personal) POI database $\mathcal{P}_{p+c}$ and a session $S$,
the personalized visited-POI assignment task is to find $\{ (sp_n,poi_n) \}_{n=1}^M$ under a condition of $sp_{n_1} \cap sp_{n_2} = \emptyset$ for all $n_1,n_2 \in\{1,\dots, M\}$ and $n_1\neq n_2$,
where
$poi_n \in \mathcal{P}_{p+c}$, and $sp_n \subseteq S$.
Moreover,  $M$ represents the number of true visited-POIs obtained from given $(\mathcal{P}_{p+c}, S)$.
Therefore, the input and output of our task are $\mathcal{I}=(\mathcal{P}_{p+c}, S)$ and $\mathcal{O} = \{ (sp_n,poi_n) \}_{n=1}^M$, respectively.

\section{Related Work}
\label{sec:related_work}
This section explains related work in terms of several perspectives of our proposal.

\subsection{GPS trajectory mining and stay-point extraction}
\label{sec:stay_point}
Many studies on GPS trajectory mining exist, such as user activity estimation \cite{Cho:2011,Farrahi:2011,Liao:2005a,Liao:2007}, transportation mode detection \cite{Zheng:2008a,Zheng:2011}, and region analysis \cite{Xiao:2010,Yuan:2012a}.
A typical approach to tackle these tasks first extracts stay-points as a clue for solving them.
Therefore, we believe that stay-point extraction is a key technology of many GPS trajectory mining tasks.

Various stay-point extraction methods have already been proposed.
For example, Ashbrook and Starner~\cite{Ashbrook:2002,Ashbrook:2003} use a modified $k$-means method, Adams et al.~\cite{Adams:2006} use DBSCAN~\cite{Ester:1996}, and Kurashima et al.~\cite{Kurashima:2010} employ Mean-Shift~\cite{Cheng:1995}, all of which are based on clustering.
Kang et al. \cite{Kang:2004} and Zheng et al. \cite{Zheng:2009} assume that stay-points are positions within a constant radius from a center where the stay time exceeds a constant time.
More recently, Nishida et al. \cite{Nishida:2015jc} developed a more robust stay-point extraction algorithm that considers outliers and missing points in GPS trajectories.

As in other GPS trajectory mining tasks,
we can leverage a stay-point extraction method to obtain the clues of visited-POIs in our visited-POI assignment task.
However, it is nearly impossible to extract only meaningful stay-points for the downstream task that we are actually aiming to solve (visited-POI assignments) by simply applying a conventional stay-point method.
This is because the above stay-point extraction methods were developed independently of downstream tasks, and thus they extract stay-points regardless of the objectives of the downstream tasks.
Thus, treating all the identified stay-points as if they were all visited-POIs does not serve the purpose of our task.

\subsection{Detecting semantic locations}
The challenge that most resembles our personalized visited-POI assignment task is detecting semantic locations from GPS trajectory data \cite{Cao:2010,Liu:2006,Zheng:2009}.
Cao et al. \cite{Cao:2010} extracted stay-points from trajectories and combined them with street addresses obtained by a reverse geocoder.
Their method assigns a semantic label to stay-points by yellow-pages.
This strategy resembles a nearest neighbor assignment to stay-points by a POI database.
Liu et al. \cite{Liu:2006} also extracted semantic locations from GPS trajectories in the same manner.
Zheng et al. \cite{Zheng:2009} extracted stay-points from user trajectories and applied a hierarchical clustering algorithm to combine stay-points to create hierarchical stay areas on a diagram called a tree-based hierarchical graph.
The key difference between our personalized visited-POI assignment task and a semantic location detection task is that the semantic location is essentially determined on the basis of stay-points, while in this paper we determine a visited-POI on the basis of whether the user actually visits it.

\subsection{POI recommendations}
POI recommendation tasks are closely related to our target task.
Many previous studies have addressed POI recommendations~\cite{Bao:2012,Leung:2011,Wang:2013,Ye:2011,Yuan:2013}.
Most used the collaborative filtering (CF) approach, which requires inter-user information, to achieve recommendations.
Leung et al. \cite{Leung:2011} performed co-clustering on users and stay-points to improve CF recommendations.
Ye et al. \cite{Ye:2011} proposed a framework that fuses user preferences to a POI with both social and geographical influences.
Wang et al. \cite{Wang:2013} showed that the most frequently used check-in history in location-based social networks is first-visit POIs and
proposed a personalized PageRank-based method to improve the accuracy of estimating first-visit POI recommendations.
Yuan et al. \cite{Yuan:2013} introduced a time-aware feature into a CF-based approach and
showed that incorporating temporal and spatial influences improves the accuracy of POI recommendations.
These studies exploit other user check-in histories to recommend POIs to users.
Bao et al. \cite{Bao:2012} studied location recommendation with a location category hierarchy
and concluded that since different users have varying levels of expertise and preferences, they should be treated differently in the recommendation process.

These CF-based methods used in the POI recommendation task resemble our method.
However, they require the histories of many users; our target task, which is a personalized visited-POI assignment, only assumes using individual GPS trajectories and the annotation data of users.
Therefore, since these methods are related without being direct competitors of our proposed method, we ignore their evaluations in our experiments.

Other studies on a location naming task \cite{Lian:2011,Lian:2014} and a POI recommendation task \cite{Shaw:2013} use a supervised learning algorithm \cite{Cao:2007b,Liu:2011} to build POI ranking models.
To formalize their problems as a ranking challenge, they look at a location (i.e., longitude and latitude) as a query and a user's check-in data to it as relevant labels.
Their methods utilize user history, the statistics of POIs in a check-in service, and other information to generate features.
Then the ranking model uses the features to rank POI candidates.
The key difference between a visited-POI assignment task and a POI recommendation task is the latter's requirement for significant location extraction.
Previous POI recommendation studies assume that significant locations are given, but our visited-POI assignment task does not.
One straightforward approach is cascading a stay-point extraction algorithm and a POI recommendation method.
We regard the nearest neighbor method \cite{Cao:2010} and learning-to-rank methods \cite{Lian:2011,Shaw:2013} as similar approaches to our proposed method\footnote{We do not consider the method by Lian et al. \cite{Lian:2014} to be a similar method because it requires check-in histories of many users to calculate latent topic features.  Its other part is equivalent to another previous work \cite{Lian:2011}.}.

\subsection{Other related tasks}
More recently, several novel tasks have been proposed that are related to visited-POI assignments.
For example,
Lu et al. \cite{Lu:2016:Ubicomp} characterized the life cycle of POIs
and investigated the POI evolution process over time.
Espin-Noboa et al. \cite{EspinNoboa:2016:WWW} tackled a task that clusters users based on spatio-temporal dimensions with a non-negative tensor factorization method.
They clearly focus on different targets.

Zarezade et al. \cite{Zarezade:2016:CoRR} focused on the periodic behaviors of users and formalized POI check-in patterns as a stochastic point process.
An interesting aspect of their method is that they take into account a factor of the influence of the close friends of users.
In contrast, our task detects actual visited-POIs from obtained raw GPS trajectories and POI information, which includes the user's periodic behaviors without being limited to them.
Therefore, our task indirectly includes their task, even though it does not specifically focus on periodic behaviors.

Wang et al. \cite{Wang:2016:ICDE} proposed a sequential personalized spatial item recommendation framework (SPORE), which
recommends a sequence of POIs based on individual POI-visit histories.
Their target closely resembles ours.
However, the essential difference is that their task assumes a sequence of check-in records as input, unlike raw GPS trajectories for our case.
This means that their method does not assume that an input sequence (check-in records) contains any false positive information, which is one of the main challenges of our task.
In addition, SPORE, their proposed algorithm, cannot be directly applied to GPS trajectories since it does not have a mechanism that removes false positive stay-points, while our method can remove such meaningless stay-points.

Lv et al.~\cite{Lv:2012:DPS:2396761.2398471} proposed a task that detects personally semantic places from GPS trajectories.
Their proposed task also appears to closely resemble ours.
However, their target is to detect places ({\it frequently} visited by an individual user) that might have such important semantic meanings as home or office.
In this perspective, their target is closely related to Zarezade et al. \cite{Zarezade:2016:CoRR}, as explained above.
In contrast, our proposed task detects not only frequently visited places like homes and offices but also every POI that the user actually visits regardless of the frequency.

Keles et al.~\cite{Keles:2017:EVP:3080546.3080552} employed a Bayesian network to detect the categories of visited-POIs, such as hospitals and universities, from the GPS trajectories of vehicles.
Their motivation is closely related to ours.
The essential difference is that they only detect the categories of visited-POIs; we detect the visited-POIs themselves.
Additionally, they used vehicles' GPS trajectories, whereas we target the trajectories obtained from the mobile devices of users.
Thus, our challenge is much more complicated.

As discussed above, motivation, goals, and task settings of all the studies differ from our task even though all of these proposed tasks are related.

\section{Joint Estimation of Significant Locations and Their Visited-POIs}
\label{sec:proposed_method}
In this section, we describe how we model and solve the proposed task, personalized visited-POI assignment task.
We first discuss the properties that our model must provide and suggest a mathematical formulation that implements our proposal.
We also elaborate on the features needed to capture user behaviors.
Finally, we explain the procedure through which our model simultaneously extracts a significant location and its visited-POI from raw GPS trajectories.

\begin{figure}[t]
\centering
\includegraphics[width=13cm, clip]{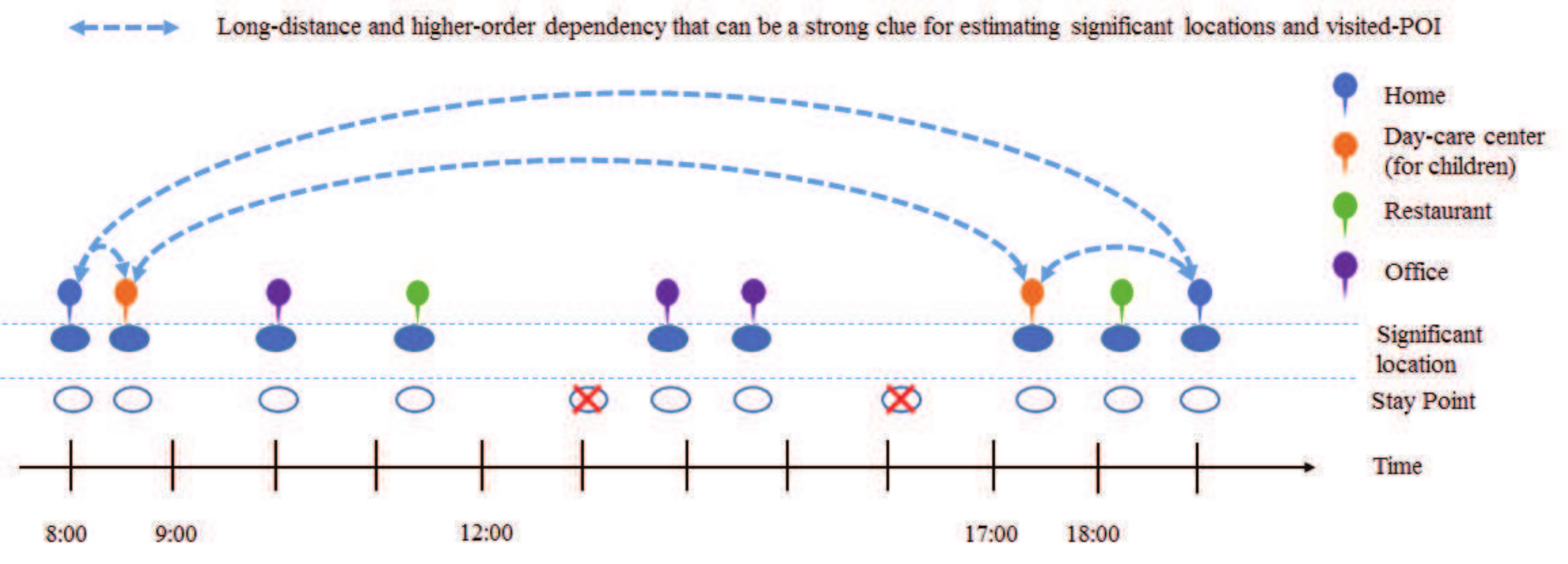}
\caption{Example of long-distance and higher-order dependency}
\label{fig:long_distance_dependency}
\end{figure}

\subsection{Two-step approach: Simplification for practical modeling}
It is computationally infeasible to consider all the durations of the track-points as candidates of significant location for assigning visited-POIs.
To avoid such computational difficulty,
we utilize a two-step approach to derive an effective algorithm for personalized visited-POI assignments.
The following is a brief explanation of our two-step procedure:
\begin{enumerate}
\item Extract stay-points using stay-point extraction with a high-recall setting, and collect the visited-POI candidates for each extracted stay-point.
\item Evaluate whether each extracted stay-point is a visited-POI.
\end{enumerate}
As we described in Section~\ref{sec:stay_point}, many previous methods applied to GPS trajectory mining tasks leverage a stay-point extraction method as a first-step process to obtain clues about the target mining tasks.
We also follow this approach for our personalized visited-POI assignment task.
Note that the stay-point extraction method is not limited to any specific scheme.

\subsection{Required properties for modeling}

The remaining part of our task is selecting significant locations from stay-points and assigning a visited-POI to each one.
We discuss the requirements for an ideal method of selecting significant locations and assigning a visited-POI to them.
In our approach, the method must be able to detect a significant location from an exhaustive amount of stay-points.
For example, as explained in Section~\ref{sec:problem_setting}, the method must determine whether each stay-point is meaningless (such as traffic jams) or a significant location.
To successfully do so, we need to simultaneously consider different aspects to accurately solve our task,
for example, estimating the relation between a significant location and its visited-POI.
We also need to satisfy the consistency of the significant location of each estimated visited-POI.
This is because the significant location of any two visited-POIs never overlaps,
and such higher level knowledge as POI-POI interactions might provide very useful information for solving our task.
According to the above analyses, a Hidden Markov Model (HMM) \cite{Murphy:2012} and a Conditional Random Field (CRF) \cite{Murphy:2012} are possible candidates for modeling the personalized visited-POI assignment task.
This is because a $k$-th order Markov model (HMM/CRF) can capture $k$-consecutive patterns of visits and is efficiently solvable through dynamic programming (DP).

However, the task's essential information is often retained in the long-distance (non-consecutive) and higher-order dependency information, which cannot be represented by $k$-consecutive patterns of visits.
For example, consider a situation where a father drops off his daughter at day-care on his way to work in the morning, moves around various offices during the day, and finally picks her up after work in the evening.
Fig.~\ref{fig:long_distance_dependency} illustrates this example.
In this example, if the father drops his daughter off at day-care on his way to work in the morning, then he must pick her up after work on his way back home with very high probability.
In contrast, he does not have to visit the day-care before returning home if he did not take his daughter to day-care in the morning.
Therefore, the following order of four POIs can be a visiting-pattern block for him: (home)-(day-care)-(day-care)-(home).
The important point here is that this pattern doesn't have to be consecutive; he might (randomly) visit other places, such as a restaurant, a drug store or a museum, between the day-care center and his home, regardless of the pattern of visits.
This implies that a user's subsequent visits do not always depend only on the last (consecutive) visited-POI; they often depend on the visited-POIs long before the user visited them.
Thus, we aim to include all the combinatorial dependencies among all the stay-points and the POIs.
However, there is dilemma that such long-distance and higher-order relations cannot be efficiently calculated by the DP in the $k$-th order Markov model (HMM/CRF) in practical $k$.
In fact, DP's calculation cost may become identical as that of a naive exhaustive search if we consider all of the long-distance and higher-order dependency information.
Such a naive exhaustive search is infeasible if the number of candidates (number of stay-points and POIs) is too large.

As an alternative for much better modeling than HMM/CRF,
we formulate our task as a 0-1 integer linear programming (0-1 ILP) problem.
An ILP is a good choice to efficiently solve exhaustive searches or even NP-hard problems.
A general ILP solver, which is a package that aggregates many mathematical optimization techniques,
discards non-optimal solutions to reduce the search cost based on the constraints of an ILP problem.
The search cost depends on a problem's difficulty.
In this paper, we verify that our problem can be solved by a general ILP solver in a reasonable time.
ILP formalization is a general form that embraces both HMM and CRF \cite{Roth:2005}.
Since a special case of ILP formalization perfectly conforms to HMM/CRF,
we emphasize that our framework includes HMM/CRF.

\begin{table}[t]
\centering
\caption{Notation for our formulation\label{table:notation}}{%
\begin{tabular}{l|p{120mm}}\hline
 Symbol & Description \\\hline\hline
 $s_i, \bar{s}_i $ & Pair of variables that represent whether the $i$-th stay-point is a significant location ($s_i=1$, $\bar{s}_i=0$) or not ($s_i = 0$, $\bar{s}_i=1$).\\\hline
 $v_{ik} $ & Variable that represents whether the $k$-th visited-POI candidate at the $i$-th stay-point is a visited-POI ($v_{ik}=1$) or not ($v_{ik}=0$). \\\hline
 $t_{ijkl} $ & Variable that represents whether the $k$-th visited-POI candidate in the $i$-th stay-point and the $l$-th visited-POI candidate in the $j$-th stay-point are both visited-POIs ($t_{ijkl}=1$) or not ($t_{ijkl}=0$). \\\hline
 $y_m $ & Variable that represents whether the number of visited-POIs is $m$ ($y_m=1$) or not ($y_m=0$). \\\hline
\end{tabular}
}
\end{table}
\begin{figure}[t]
\centering
\includegraphics[width=6cm, clip]{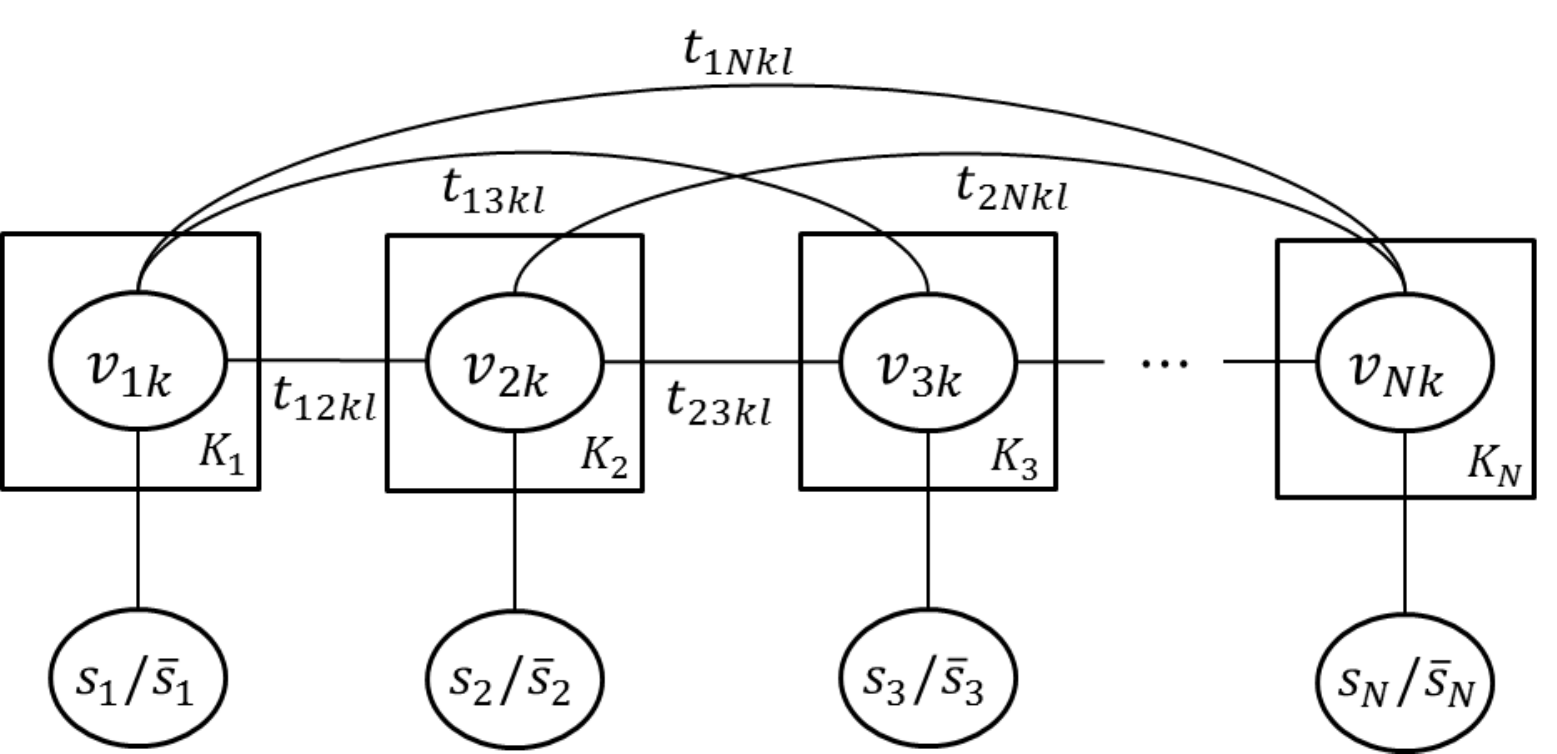}
\caption{Diagrammatic illustration of variables in our method}
\label{fig:joint_estimation_image}
\end{figure}
\subsection{Modeling by 0-1 ILP formulation}

Table \ref{table:notation} summarizes all the notation symbols and their descriptions for our formulation.
All are binary (0-1) variables.
$s_i$ and $\bar{s}_i$, which are assigned to each stay-point, denote variables that indicate whether a location is significant.
$(s_i = 1, \bar{s}_i = 0)$ if the $i$-th stay-point is a significant location, and $(s_i = 0, \bar{s}_i=1)$ otherwise.
$v_{ik}$ represents a variable for visited-POIs assigned to each visited-POI candidate.
$v_{ik}=1$, if the $k$-th POI is the visited-POI of the $i$-th stay-point, and $v_{ik}=0$ otherwise.
$t_{ijkl}$ denotes the interaction of two visited-POIs with their stay-points.
$t_{ijkl}=1$, if the $k$-th POI is the visited-POI of the $i$-th stay-point, and the $l$-th POI is the visited-POI of the $j$-th stay-point, and $t_{ijkl}=0$ otherwise.
$y_m$ also represents the visited-POI sequence length.
$y_m = 1$, if the sequence length is $m$, and $y_m=0$ otherwise.

Fig. \ref{fig:joint_estimation_image} illustrates the formalization as an 0-1 ILP problem.
A lower circle node denotes an extracted stay-point, and an upper circle node surrounded by a rectangle (plate) denotes a candidate of the corresponding visited-POI.
A plate represents node replicates, and $K_i$ denotes the number of visited-POI candidates at the $i$-th stay-point.
We also used visited-POI interaction variable $t_{ijkl}$ that corresponds to the combination of the $k$-th visited-POI candidate at the $i$-th stay-point and the $l$-th visited-POI candidate at the $j$-th stay-point.

Let ${\mathcal S}$, $\bar{\mathcal S}$, ${\mathcal V}$, ${\mathcal T}$, and ${\mathcal Y}$ respectively represent a set of all the variables of $s_i$, $\bar{s}_i$, $v_{ik}$, $t_{ijkl}$, and $y_m$.
We formalize the joint estimation task as 0-1 ILP problem $P$ as follows:
\begin{eqnarray}
P: \max_{\mathcal{S}, \bar{\mathcal{S}}, \mathcal{V}, \mathcal{T},\mathcal{Y}}
 &&
 \sum_i \left\{ \langle \mathbf{w}^{(s)},
         \mathbf{x}^{(s_i)} \rangle s_i + \langle
         \mathbf{w}^{(\bar{s})},
         \mathbf{x}^{(\bar{s}_i)} \rangle \bar{s}_i
        \right\} \label{eq:1} \\
&& + \sum_i \sum_k \langle \mathbf{w}^{(v)},
 \mathbf{x}^{(v_{ik})} \rangle v_{ik} \label{eq:2} \\
&& + \sum_i \sum_j \sum_k \sum_l \langle
 \mathbf{w}^{(t)}, \mathbf{x}^{(t_{ijkl})} \rangle
 t_{ijkl}\label{eq:3} \\
&&  + \sum_m \langle \mathbf{w}^{(y)}, \mathbf{x}^{(y_m)}
 \rangle y_m \label{eq:4}  \\
\text{subject to: }
 &&
 s_i + \bar{s}_i = 1, \hspace{0.2em} \sum_{(i,j) \in L} s_i + s_j \le 1 \label{eq:5} \hspace{0.5em} \forall i\\
&&  \sum_k v_{ik} = s_i \hspace{0.5em} \forall i \label{eq:5.1} \\
&& \sum_k \sum_l t_{ijkl} \le 1  \hspace{0.5em} \forall i,j
\label{eq:6} \\
&& t_{ijkl} \le v_{ik}, \hspace{0.3em} t_{ijkl} \le v_{jl}
\hspace{0.5em} \forall i,j,k,l \label{eq:7} \\
&& \sum_m y_m = 1, \hspace{0.3em} \sum_m m y_m = \sum_i
s_i.
\label{eq:8}
\end{eqnarray}

Here, $\mathbf{x}^{(\cdot)}$ denotes a feature vector that corresponds to variable $(\cdot)$, and weight vector $\mathbf{w}^{(\cdot)}$ denotes a weight vector for feature $\mathbf{x}^{(\cdot)}$.
$\langle \cdot, \cdot \rangle$ denotes the inner product of the given vectors.
In Eq. (\ref{eq:5}), $L = \{(i,j)|i,j\in I, i \neq j, {\rm ol}(sp_i, sp_j)\}$, where $I$ denotes the set of stay-point indexes, and ${\rm ol}(sp_i, sp_j)$ returns true if stay-points $i$ and $j$ overlap.

The objective function consists of the following four parts: (1) stay-point features, (2) stay-point and POI features, (3) POI-POI features, and (4) sequence length features.
Each feature value is calculated using a user's annotated data.
Weight vectors $\mathbf{w}^{(\cdot)}$ are also estimated from the user's annotated data.
Note here that training the features and the weights are the personalized parts in our approach since we basically train them just using a single user's personal data.
We expect that such training can capture some personalized behaviors in the features and the weights.

After we describe the details and the calculation methods of these features, we explain the details of the constraints of problem $P$.

\subsection{Features and constraints in ILP formulation}

This section describes the features and constraints in Eqs. (\ref{eq:1}) to (\ref{eq:8}).

\subsubsection{Stay-point features}
Stay-point features $\mathbf{x}^{(s_i)}$ and $\mathbf{x}^{(\bar{s}_i)}$ in Eq. (\ref{eq:1}) show the validity and the invalidity of stay-point $i$ as a significant location.
Higher values in $\mathbf{x}^{(s_i)}$ indicate that it is more reasonable to choose stay-point $i$ as a significant location ($s_i=1$).
In this paper, we prepare two features, $\mathbf{x}^{(s_i)} = (x^{(s_i)}_1, x^{(s_i)}_2 )$, to capture the validity of a stay-point.
As for $\mathbf{x}^{(\bar{s}_i)} = (x^{(\bar{s}_i)}_1, x^{(\bar{s}_i)}_2)$, we use $x^{(\bar{s}_i)}_1 = 1 - x^{(s_i)}_1$ and $x^{(\bar{s}_i)}_2 = 1 - x^{(s_i)}_2$ to create features.

For $x^{(s_i)}_1$, we use a Gaussian basis function to utilize a user's significant location history information.
This function calculates the value based on the center of stay-point $i$ and the nearest stay-point in the annotated data in the following equation:
\begin{equation}
x^{(s_i)}_1 = \exp \left\{ - \gamma \parallel sp_i - \mathrm{NN}_{\rm train}(sp_i) \parallel_2^2 \right\}.\nonumber
\end{equation}
Here, $\gamma$ is an accuracy parameter of the Gaussian basis function, which controls the decreasing degree of the function value.
If $\gamma$ has a large value, it decreases more rapidly with distance, and if it has a small value, it decreases less rapidly.
$\mathrm{NN}_{\rm train}(\cdot)$ returns a stored significant location whose center is the nearest to the input stay-point.
$\parallel sp_i - sp_j \parallel_2$ calculates the geographical distance between $sp_i$ and $sp_j$.

As a second feature, $x_2^{(s_i)}$, we model the validity of stay-point $i$ based on its stay time to capture the natural characteristic that a longer time stay implies a higher probability as a significant location.
We use the cumulative probability function of the exponential distribution to calculate the feature values.
The distribution describes the time between events in a process in which events occur continuously and independently at a constant average rate.
The cumulative probability function is calculated by
\begin{equation}
 x_2^{(s_i)} = 1-\exp(- \lambda sp_i.st ),\nonumber
\end{equation}
where $\lambda$ denotes a parameter of the average event number per unit time.
The parameter can be manually fixed (e.g., $\lambda=1/30$ if the average interval time is $30$ minutes) or estimated using individually annotated data.

\subsubsection{Stay-point and POI-interaction features}
\label{sec:sp_poi_feature}
The second part of the objective function represents the validity of visited-POI $k$ at stay-point $i$.
Suppose each POI generally has a {\it POI category} that is preliminarily defined by a thesaurus, such as home, train station, or office.
The POI category plays a critical role when we estimate the visited-POIs for roughly capturing personal behaviors in the model.
We used three features for $\mathbf{x}^{(v_{ik})}$: (1) a feature based on a user's visited-POI histories, (2) a user's visited-POI category histories, and (3) POI-category stay-time information.

For a user's visited-POI history, we use multinomial distribution to model the historical information:
\begin{equation}\label{eq:sp_poi_feature1}
x^{(v_{ik})}_1 = \alpha_1 f^{\rm time}_{\rm poi} (k, sp_i.bt) + (1 -  \alpha_1) f_{\rm poi}(k),
\end{equation}
where $f_{\rm poi}(\cdot)$ denotes the probability of visited-POI $k$, which is estimated based on the ratio of POI $k$ in the user's annotated data.
$f^{\rm time}_{\rm poi}(\cdot,\cdot)$ denotes the time-dependent version of $f_{\rm poi}(\cdot)$: probability of visited-POI $k$ in $sp_i.bt$.
We split a day's 24 hours into four six-hour windows to calculate $f^{\rm time}_{\rm poi}(\cdot, \cdot)$.
We also prepared a feature based on POI-category information in a similar way:
\begin{equation}\label{eq:sp_poi_feature2}
x^{(v_{ik})}_2 = \alpha_2 f^{\rm time}_{\mathrm{cat}} ({\rm cat}(k), sp_i.bt) +
(1 - \alpha_2) f_{\mathrm{cat}} ({\rm cat}(k)), \nonumber
\end{equation}
where ${\rm cat}(k)$ denotes the category of POI $k$ and $f^{\rm time}_{\mathrm{cat}}(\cdot, \cdot)$ and $f_{\mathrm{cat}}(\cdot)$ are the category versions of the functions in Eq. (\ref{eq:sp_poi_feature1}).
Here $\alpha_1, \alpha_2 \in [0,1]$ are the smoothing parameters for the linear interpolation of the time independent probability functions.

The third feature, $x^{(v_{ik})}_3$, is calculated using the stay-time likelihood of the POI category.
We assume that each POI category has its own particular probabilistic stay-time distribution, which generates the specific stay time of the visited-POIs.
For example, it takes a short time to buy a beverage at a store and a long time to have dinner at a restaurant.
We calculate the likelihood of a given stay-point and a visited-POI candidate using the probabilistic density function of the stay-time distribution.
We adopt log-normal distribution as the stay-time distribution since it is a common choice when modeling the distribution of time spent in user activities \cite{Kim:2006,Minder:2005}.
We use maximum likelihood estimation (MLE) to calculate the distribution parameters from the annotated data.

The MLE values for the log-normal distribution parameters of category $c$ are calculated by
\begin{equation}
\nu_c = \frac{1}{|I_c|} \sum_{i \in I_c} \ln (sp_i.st), \hspace{1em}
\tau_c = \frac{1}{|I_c|} \sum_{i \in I_c } \left\{ \ln (sp_i.st) -
				     \tau_c \right\}^2. \nonumber
\end{equation}
Here $I_c \equiv \{i| i \in I, {\rm cat}(sp_i.poi)=c\}$ denotes the index set of the stay-points with corresponding visited-POIs whose category is $c$.
$|I_c|$ denotes the size of set $I_c$, which is the total number of visits to POI-category $c$.
Using these parameters, we calculate feature value $x^{(v_{ik})}_3$:
\begin{equation}
x^{(v_{ik})}_3 = \mathcal{LN}(sp_i.st; \nu_{{\rm cat}(k)}, \tau_{{\rm cat}(k)}),\nonumber
\end{equation}
where $\mathcal{LN}$ is the probability density function of the log-normal distribution.

\subsubsection{POI-POI interaction features}
\label{sec:poi_poi_feature}
The third part of the objective function calculates the validity in terms of the visited-POI combinations.
This factor helps model the consistency of the visited-POIs in a session because all of the pairs of the visited-POI candidates are considered.
This aspect is also a key difference between our method and the conventional methods.
In this paper, we prepared two POI-POI features: (1) a probability of POI-POI category dependency and (2) a Jaccard coefficient between two POI categories.

The first feature, $x^{(t_{ijkl})}_1$, is calculated using the probability of two POI categories:
\begin{equation}
x^{(t_{ijkl})}_1 = P({\rm cat}(l)|{\rm cat}(k)) = \frac{{\rm num}({\rm cat}(k) \to
 {\rm cat}(l)) + \beta}{{\rm num}({\rm cat}(k)) + \beta |C|}, \nonumber
\end{equation}
where ${\rm num}({\rm cat}(k))$ denotes the number of visits to ${\rm cat}(k)$ in the annotated data, ${\rm num}({\rm cat}(k) \rightarrow {\rm cat}(l))$ denotes the number of visits to ${\rm cat}(l)$ after a visit to ${\rm cat}(k)$, $\beta$ denotes the smoothing parameter, and $|C|$ denotes the total number of categories.

As a second feature, we use a Jaccard coefficient, a well-known technique in data mining, to capture the connection between two factors.
We calculate the Jaccard coefficient between two POI categories by
\begin{equation}
x^{(t_{ijkl})}_2 = {\rm Jaccard}({\rm cat}(k), {\rm cat}(l)) = \frac{|S( {\rm cat}(k) \cap {\rm cat}(l))|}{|S({\rm cat}(k))| + |S({\rm cat}(l))|}, \nonumber
\end{equation}
where $S({\rm cat}(k) \cap {\rm cat}(l))$ denotes a set of sessions containing both categories ${\rm cat}(k)$ and ${\rm cat}(l)$.

\subsubsection{Sequence length features}
This part plays the role of a regularizer that controls the number of visited-POIs in a session.
That is, it enables the objective function to avoid assigning too many or too few visited-POIs.
We prepared binary variable $y_m$ that takes 1 if the number of visited-POIs is $m$ and 0 otherwise.

We assume that the validity of the visited-POI number can be modeled by Gaussian distribution and
use its probability density function to create feature $x^{(y_m)}_1$: $x^{(y_m)}_1 = \mathcal{N}(m; \mu_y, \sigma_y^2)$, where $\mu_y$ and $\sigma_y^2$ are the mean and variance parameters.
Each user has his/her own such parameters, which we estimate on the basis of MLE from the user's annotated data.

\subsubsection{Constraints}
Each constraint in problem $P$ plays the important role of satisfying the requirements for the visited-POI assignment task.
For example, $s_i=1$ and $v_{ik}=0 \hspace{0.2em} \forall k$ are supposedly impermissible results.
In this example, $s_i=1$ indicates that stay-point $i$ is a significant location, and $v_{ik}=0 \hspace{0.2em} \forall k$ indicates there is no visited-POI in stay-point $i$.
These results contradict our assumption.
To avoid such problems, we introduce our assumptions into the problem as constraints and
prepare the following:
\begin{itemize}
\item Eq. (\ref{eq:5}): The left constraint ensures that stay-point $i$ is either a significant location or a non-significant location. The right constraint ensures the exclusive selection of overlapped stay-points.
\item Eq. (\ref{eq:5.1}): If stay-point $i$ is a significant location, there is one corresponding visited-POI.
\item Eq. (\ref{eq:6}): There is only one dependency pattern from stay-points $i$ to $j$.
\item Eq. (\ref{eq:7}): If $t_{ijkl}=1$, visited-POI at stay-point $i$ is POI $k$, and visited-POI at stay-point $j$ is POI $l$.
\item Eq. (\ref{eq:8}): A session length, which is exactly a single value (left constraint), is calculated by the sum of the adopted significant locations (right constraint).
\end{itemize}

Note that the exclusive selection factor in Eq. (\ref{eq:5}) plays a particularly important role when different algorithms are used to extract stay-points because some cannot be adopted as significant locations.

\subsection{Algorithm}
We show our algorithm in Fig. \ref{fig:algorithm}.
It receives a set of stay-points in a session as given significant location candidates and weight parameters.
For each significant location candidate, its visited-POI candidates are obtained by a POI database $\mathcal{P}_{p+c}$.
It also creates features for each stay-point and a corresponding visited-POI candidate and
generates an ILP problem on the basis of the calculated feature and the given weight parameters.
It then solves the problem using a general ILP solver.
Finally, the algorithm returns significant locations and their visited-POIs $\mathbf{V}$ as results.

\begin{figure}[t]
\centering
\begin{tabular}{l}\hline
\textbf{Algorithm} Joint Estimation of Significant Location \\ and its Visited-POI \\\hline
\textbf{Input}: $SP = (sp_i)_{i=1}^N$, ${\bf w}^{(s)}, {\bf w}^{(\bar{s})}, {\bf w}^{(v)}, {\bf w}^{(t)}, {\bf w}^{(y)}$ \\
\textbf{Output}: $\mathbf{V}$ \\
~1: \textbf{FOR} $i$ in 1 to $N$ \\
~2: \hspace{1.0em} Create features $\mathbf{x}^{(s_i)}$ and $\mathbf{x}^{(\bar{s}_i)}$.\\
~3: \hspace{1.0em} Obtain POI candidates $\{poi_k \}_{k=1}^{K_i}$. \\
~4: \hspace{1.0em} \textbf{FOR} $k$ in 1 to $K_i$ \\
~5: \hspace{2.0em} Create feature $\mathbf{x}^{(v_{ik})}$.\\
~6: \hspace{1.0em} \textbf{ENDFOR}\\
~7: \textbf{ENDFOR}\\
~8: Create feature $\mathbf{x}^{(t_{ijkl})}$.\\
~9: Create feature $\mathbf{x}^{(y_m)}$.\\
10: Generate 0-1 ILP code and solve the problem. \\
11: Read $\mathbf{V}$ from the solution. \\
12: \textbf{RETURN} $\mathbf{V}$\\ \hline
\end{tabular}
\caption{Algorithm of our framework}
\label{fig:algorithm}
\end{figure}

\subsection{Computational complexity}
It is computationally expensive to solve the personalized visited-POI estimation task with incorporating all the long-distance dependencies by a naive exhaustive search since we need to consider all the combinatorial dependencies among all the estimated significant locations and their visited-POIs.
However, a recent-advanced general purpose ILP solver (e.g., Gurobi, and CPLEX) can solve many ILP problems in a practical time even if a given problem is NP-hard since it equips various strong techniques for efficiently solve sophisticated combinatorial problems.
In fact, the number of variables in our problem is relatively very small in terms of a standard problem size handled currently in generic ILP solvers,
and thus, the execution time for estimating visited-POIs from a single session (one day's data) by our method generally takes less than one second (see details in experimental results).

Note that we can model long-distance dependencies by the $k$-th order Markov model with a sufficiently large $k$.
Theoretically, its calculation cost is a polynomial time if we apply an appropriate dynamic programming algorithm.
However, it actually becomes a large $k$-th polynomial time, for example $k=20$, which is basically infeasible.

\begin{table}[t]
\centering
\caption{Statistics for annotated dataset. (\#Sessions: number of sessions. \#POIs: number of POIs. Ave. POI/Session: average number of POIs per session. \#Uniq. POIs: number of unique POIs per user. \#count=1 POIs: number of POIs appeared only once per user. \#Uniq. POI cate.: number of unique POI categories.) \label{table:annotated_dataset_stats}}{
\tabcolsep=2pt
\begin{tabular}{c|rrrrrr}\hline
User & \#Sessions & \#POIs & (Ave. POI/Session)& \#Uniq. POIs &(\#count=1 POIs)& \#Uniq. POI cate. \\ \hline\hline
A01	&	16	&	98	&	(6.13) 	&	24	&	(15)	&	18	\\
A02	&	20	&	89	&	(4.45) 	&	43	&	(31)	&	28	\\
A03	&	18	&	95	&	(5.28) 	&	43	&	(35)	&	32	\\\hline
B04	&	21	&	119	&	(5.67) 	&	56	&	(46)	&	32\\
B05	&	16	&	112	&	(7.00) 	&	41	&	(31)	&	26\\
B06	&	18	&	116	&	(6.44) 	&	60	&	(50)	&	37\\
B07	&	19	&	99	&	(5.21) 	&	47	&	(40)	&	27\\
B08	&	19	&	115	&	(6.05) 	&	35	&	(26)	&	19\\
B09	&	20	&	221	&	(11.05) &	119	&	(97)	&	36\\
B10	&	19	&	120	&	(6.32) 	&	41	&	(26)	&	21\\
B11	&	20	&	168	&	(8.40) 	&	92	&	(77)	&	39\\
B12	&	17	&	162	&	(9.53) 	&	43	&	(32)	&	21\\
B13	&	20	&	159	&	(7.95) 	&	84	&	(62)	&	52\\
B14	&	10	&	67	&	(6.70) 	&	38	&	(33)	&	30\\
B15	&	17	&	87	&	(5.12) 	&	39	&	(33)	&	28\\
B16	&	21	&	85	&	(4.05) 	&	16	&	(11)	&	13\\
B17	&	19	&	85	&	(4.47) 	&	30	&	(21)	&	18\\
B18	&	19	&	87	&	(4.58) 	&	41	&	(37)	&	20\\
B19	&	21	&	95	&	(4.52) 	&	39	&	(32)	&	20\\
B20	&	16	&	74	&	(4.63) 	&	40	&	(38)	&	31\\
B21	&	21	&	88	&	(4.19) 	&	20	&	(11)	&	11\\\hline\hline
ALL	&	387	&	2341	&	(6.05) 	&	971	&	(758)	&	202	\\\hline
\end{tabular}
}
\end{table}

\section{Evaluation}
\label{sec:evaluation}
 \subsection{Data}
 \label{sec:data}
 We collected the actual GPS trajectories and the corresponding visited-POI annotations of users to evaluate our proposed task: personalized visited-POI assignments\footnote{Unfortunately, no publicly available dataset contains both individual GPS trajectories and user's visited-location information.}.
First, we collected the trajectories of three subjects (users) in the first round of our challenge for preliminary evaluations and obtained 18 more from the second round.
Thus, we collected a total of 21 subject trajectories and corresponding visited-POI annotations.
Each subject carried a mobile device with a logging application for three weeks (21 days) for both the first and second rounds \footnote{In
Table \ref{table:annotated_dataset_stats},
some subject trajectories were actually less than 21 sessions (days) for several reasons, e.g., they didn't go anywhere or didn't visit any significant locations (mostly on weekends), forgot to turn on (or bring) the system, or privacy issues.
}.
The logging application stores position information every three seconds with an Android OS location class that uses both GPS and WiFi positioning systems.
The application automatically adopts the position with a higher accuracy value.
Subjects also assigned visited-POIs to their significant locations.
The visited-POI candidates were collected by Foursquare Search Venues API\footnote{\url{https://developer.foursquare.com/docs/venues/search}}.
If a true visited-POI did not appear in the collected POI candidates, then the subject added custom POIs.
Table \ref{table:annotated_dataset_stats} shows the fundamental statistics of our dataset.
The numbers from 01 to 21 in the first column indicate the subject IDs.
Single letters `A' and `B' of the prefixes of the subject IDs represent one of two distinct time periods during which the data were actually collected (A for the first round and B for the second).
We collected almost 400 sessions and over 2,300 actual visited-POIs, which is a relatively large amount of GPS data with actual user visiting histories.

\begin{table}[t]
\centering
\caption{Statistics of our data: distance from stay-point center to visited-POI and average ranking of true visited-POIs among visited-POI candidates based on distance\label{table:poi_rank_dist}}{
\tabcolsep 3pt
\begin{tabular}{l|rr|rrrrrr}\hline
User & Avg. distance  [m]& (longest) & Ave. rank & \#Top-rank &   (ratio)& \#Worst-rank\\\hline\hline

A01 &$  32.57\pm 15.49$&	( 79.56)&  5.74  $\pm$   8.71& 36 / 98 &	(0.3673) 	& 48     \\
A02 &$  37.00\pm 60.08$&	(298.88)& 12.80  $\pm$  24.40& 40 / 89 &	(0.4494) 	&130     \\
A03 &$  28.78\pm 37.35$&	(332.61)& 12.70  $\pm$  20.57& 19 / 95 &	(0.2000) 	&111	\\\hline

B04 &$  34.30\pm 39.89$&	(242.91)& 13.01   $\pm$14.58 & 11  /119 &	(0.0924) 	&        67     \\
B05 &$  49.04\pm 40.74$&	(171.55)& 18.17   $\pm$21.39 & 11  /112 &	(0.0982) 	&        66      \\
B06 &$  49.37\pm 49.34$&	(290.92)& 16.07   $\pm$19.85 & 22  /116 &	(0.1897) 	&        77     \\
B07 &$  33.72\pm 40.09$&	(199.38)&  9.25   $\pm$12.35 & 17  / 99 &	(0.1717) 	&        48     \\
B08 &$  37.67\pm 30.80$&	(149.07)&  9.46   $\pm$11.68 & 14  /115 &	(0.1217) 	&        61     \\
B09 &$  28.70\pm 39.08$&	(209.27)&  5.05   $\pm$ 8.76 & 73  /221 &	(0.3303) 	&        58     \\
B10 &$  40.00\pm 35.42$&	(196.41)& 11.56   $\pm$14.20 & 5   /120 &	(0.0417) 	&        71     \\
B11 &$  39.04\pm 46.33$&	(320.95)& 10.45   $\pm$15.16 & 57  /168 &	(0.3393) 	&        73     \\
B12 &$  47.46\pm 34.79$&	(187.65)& 15.68   $\pm$14.72 & 9   /162 &	(0.0556) 	&        68     \\
B13 &$  50.31\pm 72.54$&	(475.26)& 16.34   $\pm$17.03 & 18  /159 &	(0.1132) 	&        60     \\
B14 &$  26.93\pm 46.07$&       (174.84)& 10.05   $\pm$18.24 & 12  / 67 &	(0.1791) 	&        68     \\
B15 &$  29.80\pm 32.07$&	(151.67)&  8.68   $\pm$10.92 & 12  / 87 &	(0.1379) 	&        60     \\
B16 &$ 108.67\pm131.1$&       (465.27)&   3.66   $\pm$ 7.06 &  23  / 85 &	(0.2706) 	&        36     \\
B17 &$  33.22\pm 25.15$&	(102.37)& 13.14   $\pm$15.75 & 11  / 85 &	(0.1294) 	&        72     \\
B18 &$  42.22\pm 32.75$&	(125.19)& 16.60   $\pm$16.54 & 4   / 87 &	(0.0460) 	&        67     \\
B19 &$  44.11\pm 68.88$&	(380.89)& 17.65   $\pm$17.89 & 5   / 95 &	(0.0526) 	&        61     \\
B20 &$  26.63\pm 20.30$&	( 67.26)& 12.89   $\pm$13.76 & 5   / 74 &	(0.0676) 	&        47     \\
B21 &$  31.77\pm 22.25$&	(103.79)& 13.71   $\pm$14.15 & 6   / 88 &	(0.0682) 	&        48   \\  \hline \hline
ALL & $39.66 \pm 49.39$&    (475.26) & $15.40 \pm 21.20$& 410 / 2341 & (0.1751)  & 130 \\\hline
\end{tabular}
}
\end{table}

\subsection{Preliminary data analyses}
In this section, we describe our preliminary data analyses that were explained in the previous section.

\subsubsection{Analysis of visited-POIs:}
\label{subsec:sp_poi_analysis}
We obtained the visited-POI candidates near the center of the significant locations in the dataset using the Foursquare Search Venues API.
We calculated the distance from the center of a significant location to the position of each visited-POI candidate as well as
the distance's mean and standard deviation.
After sorting the POI candidates by distance in ascending order, we calculated the average rank of the visited-POIs and the ratio of the number of top-ranked POIs to the number of visited-POIs.
We show the results in Table \ref{table:poi_rank_dist}.

We confirmed that on average, the visited-POIs are mostly 100 meters or less from the center of the stay-point, and the average rank of the visited-POIs was 15.40.
These statistics support that visited-POIs mostly exist near the center of the stay-points.

Next we discuss why a visited-POI rarely appears as the nearest neighbor of a stay-point center.
We consider two reasons: (1) incorrect positioning of the stay-point center due to GPS/WiFi positioning errors and (2) the high density of POI candidates around the stay-points.
The occurrence of positioning errors on GPS loggers is a well-known problem, especially for indoor environments \cite{Zheng:2011}.
Kj{\ae}rgaard et al. \cite{Kjaergaard:2010} reported that even state-of-the-art GPS loggers suffer from positioning errors caused by the environment.
They found that less than 80\% of the track-points in their experiments were within 20 meters of the true points.
We conducted a preliminary experiment that identified that similar errors occurred for the track-points obtained outside a building.
GPS positioning errors seem inevitable.
As for the second reason (2), our study showed that each significant location has an average of 70.3 visited-POI candidates within 500 meters.
This indicates that the existence of other POIs prevents a visited-POI from being the nearest neighbor.
Thus, we have to predict visited-POIs from among all the POI candidates.
This complicates solving the visited-POI assignment task with a simple combination of stay-point extraction and a POI recommendation technique.

\subsubsection{Analysis of stay-point extraction}
\label{subsec:sp_extraction_analysis}
We investigated the functionality of one of the conventional stay-point extraction methods to verify the validity of our hypothesis in Section \ref{sec:proposed_method}.
We used this method because stay-point extraction is generally the first step for solving GPS trajectory mining tasks.
We also sought to confirm the effectiveness of the stay-point extraction approach if it were used to perform the first step in our visited-POI assignment task.

For this experiment, we selected the method developed by Kang et al. \cite{Kang:2004} since it is one of the most widely used stay-point extraction methods \cite{Zheng:2011a,Zheng:2009}.
It has two different types of threshold parameters.
One is a time threshold, which we refer to as $\theta_{\rm time}$.
The other is a distance threshold, which we refer to as $\theta_{\rm dist}$.
In most cases, these are preliminarily configured by hand.
Kang's method recognizes continuous track-points as stay-points if the continuous track-points for all the positions are within a $\theta_{\rm dist}$ meter radius over $\theta_{\rm time}$ seconds.
As an example of this method's usage, Zheng et al. used it to configure $\theta_{\rm time} = 1800$, $\theta_{\rm dist} = 200$ to extract stay-points \cite{Zheng:2011a}.

\begin{table}[t]
\centering
\caption{Evaluation of Kang's stay-point extraction algorithm using various parameter combinations on user A01's dataset. Bold font numbers indicate maximum values, and underlined numbers indicate minimum values in columns.\label{table:stay_point_extraction}}{
\begin{tabular}{cc|cc|ccc}\hline
$\theta_{\rm dist}$ & $\theta_{\rm time}$ & $PRE_{sp}$ & $REC_{sp}$ & $TP_{sp}$ & $FP_{sp}$ & $FN_{sp}$ \\\hline\hline
100 & 1800 & 0.857  & 0.367  & 36 & 6 & 62 \\
100 & 900 & 0.854  & 0.418  & 41 & 7 & 57  \\
100 & 180 & 0.541  & $\mathbf{0.867}$  & $\mathbf{85}$ & 72 &
		 $\underline{13}$ \\
200 & 1800 & $\mathbf{0.889}$  & $\underline{0.327}$  & $\underline{32}$
	 & $\underline{4}$ & $\mathbf{66}$ \\
200 & 900 & 0.870  & 0.408  & 40 & 6 & 58 \\
200 & 180 & 0.447  & 0.735  & 72 & 89 & 26 \\
500 & 1800 & $\mathbf{0.889}$  & 0.327  & $\underline{32}$ &
	     $\underline{4}$ & $\mathbf{66}$ \\
500 & 900 & 0.811  & 0.439  & 43 & 10 & 55 \\
500 & 180 & $\underline{0.380}$  & 0.612  & 60 & $\mathbf{98}$ & 38
		 \\\hline
\end{tabular}
}
\end{table}

We evaluated how Kang's method with various parameter settings performed the first step of the visited-POI assignment.
If the algorithm successfully extracted a stay-point that includes the timestamp of an annotated visited-POI, we evaluated it as correct and refer to it as a positive ($TP_{sp}$).
We evaluated the extracted stay-point as a false positive ($FP_{sp}$) if it does not include any annotated true visited-POIs.
On the other hand, we evaluated an annotated true visited-POI as a false negative ($FN_{sp}$) if the algorithm failed to extract any stay-point that includes it.
We used $TP_{sp}$, $FP_{sp}$, and $FN_{sp}$ to calculate precision $PRE_{sp}$ and recall $REC_{sp}$ of stay-point extraction with the following equation:
\begin{equation}\label{eq:precision_recall}
PRE_{sp} = \frac{TP_{sp}}{TP_{sp}+FP_{sp}}, \hspace{1em}  REC_{sp} =  \frac{TP_{sp}}{TP_{sp}+FN_{sp}}.
\end{equation}
Note that there is no chance a true visited-POI is assigned if the algorithm failed to extract any stay-point that included it.
Therefore, to achieve highly accurate visited-POI assignments, the stay-point extraction algorithm must identify the stay-points as near to the true visited-POIs as possible.

\begin{figure}[t]
\centering
\includegraphics[width=10cm, clip]{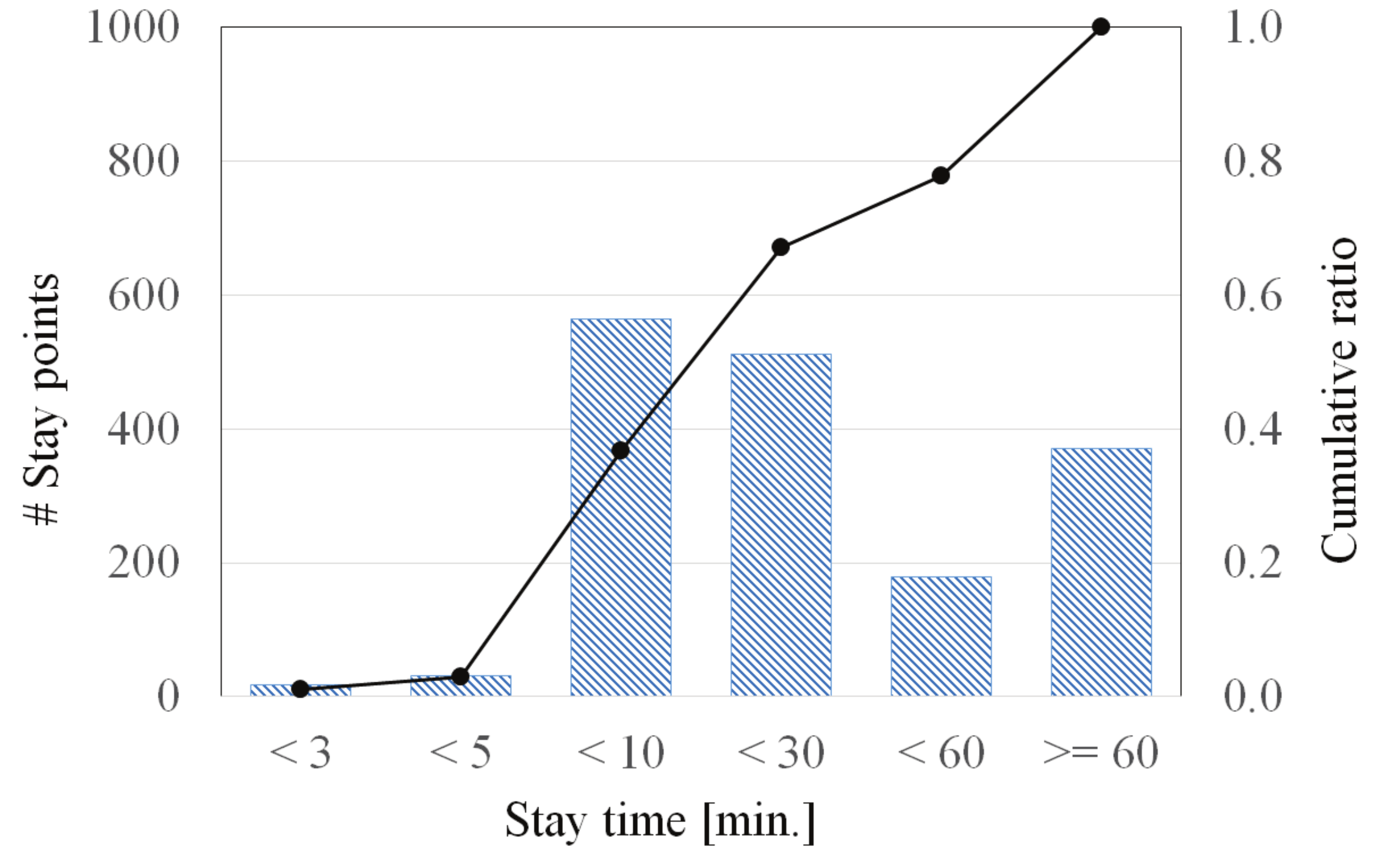}
\caption{Stay-time distribution of true significant locations in all users annotated dataset. Bars denote frequency and line denotes cumulative ratio.}
\label{fig:stay_time_distribution}
\end{figure}

Table \ref{table:stay_point_extraction} shows the number of stay-points correctly and erroneously extracted from GPS trajectories under various threshold settings.
It also shows the stay-point extraction precision and recall values obtained using Kang's method with combinations of $\theta_{\rm dist} \in \{100,200,500\}$ and $\theta_{\rm time} \in \{180,900,1800\}$.
We determined that an extracted stay-point was correct if it were less than 50 meters from its nearest true significant location.
In the table, bold font numbers denote maximum values and the underlined numbers denote minimum values in the columns.
We also analyzed the stay-time distribution of the visited-POIs in the dataset.
Fig. \ref{fig:stay_time_distribution} shows the frequency and the cumulative ratios of the stay time.

Table \ref{table:stay_point_extraction} clearly shows that with $\theta_{\rm dist}=200$, $\theta_{\rm time}=1800$ parameters, Kang's method achieved a high precision of 0.889 but a low recall of 0.327.
From Fig. \ref{fig:stay_time_distribution}, over 67\% of the stay-points had a stay time less than 30 minutes.
This indicates that with the above settings Kang's method failed to extract more than 67\% of the stay-points.
Therefore, with these settings, high recall was not achieved in our task.
On the other hand, with $\theta_{\rm dist}=100$, $\theta_{\rm time}=180$ parameters, Kang's method achieved a high recall of 0.867 while showing the lowest FN number in the column.
Although high recall performance was achieved with these settings, the precision value was low.
The most critical observation is that most true significant locations can be obtained using a conservative threshold parameter setting (e.g., $\theta_{\rm dist}=100$, $\theta_{\rm time}=180$), even though the extracted stay-points contain many false positives.

Note that the recall values in Table \ref{table:stay_point_extraction} are the upper bound recall values in the visited-POI assignment tasks.
The algorithm cannot estimate the visited-POIs if there are no extracted stay-points ($FN_{sp}$).
Similarly, the algorithm always fails to estimate the visited-POIs if the extracted stay-points are ($FP_{sp}$).
Thus, stay-point extraction errors directly lowered the accuracy of the visited-POI assignments.
This supports the validity of our proposed method's design, which first allows $FP_{sp}$ to exhaustively extract stay-point candidates and then simultaneously estimates the validity of the candidates and their visited-POIs to determine the best combination of stay-points and visited-POIs as sequences.

\subsection{Evaluation Setting}
\label{subsec:eval_visited-poi}
\subsubsection{Comparative Methods:}
We compared the following methods to verify the effectiveness of the joint estimation method.
\begin{enumerate}
\item {\bf NN}:
      The first is the Nearest Neighbor (NN) method, which selects the nearest neighbor POI from a stay-point center.
      NN resembles the visited-POI version of Cao et al. \cite{Cao:2010}.
\item {\bf NCI}:
      The second is a popularity-based method, which selects the most popular POI in terms of the number of check-ins (NCI) from neighbor POIs.
\item {\bf Rank(LI11)}:
      The third is a supervised learning-to-rank based method developed by Lian et al.~\cite{Lian:2011}.
      We used RankLib\footnote{\url{http://sourceforge.net/p/lemur/wiki/RankLib/}} as an implementation of learning-to-rank algorithms.

\item {\bf Rank(SH13)}:
      The fourth is another supervised learning-to-rank based method developed by Shaw et al.~\cite{Shaw:2013}\footnote{We removed from our implementation some of the features used by Shaw et al.~\cite{Shaw:2013} in their method because we could not obtain them from Foursquare API.}.
      We also used RankLib and Rank(LI11).

      In addition, the method proposed in Lv et al.~\cite{Lv:2012:DPS:2396761.2398471} uses a stay-point extraction algorithm and a semantic location extraction algorithm in a cascaded manner as well as our approach.
      The learning framework of their method resembles a special case of the ranking approach for Rank(LI11) and Rank(SH13) since they did not handle any POI-POI dependency information.
      Moreover, their approach basically detects frequently visited places, which is not suitable for our task.
      Therefore, we do not directly compare their method in our experiments and assume that in our task Rank(LI11) and Rank(SH13) are alternatives for Lv et al.

\item {\bf JE}:
      The fifth is the joint estimation method that considers the higher-order relation between every interaction among the information of significant locations.
      JE, which represents the full version of the proposed method in this paper,
      is our primary method.
\item {\bf CRF(JE$_{\rm 1od}$)}:
      The sixth is first-order CRF (or linear-chain CRF).
      CRF(JE$_{\rm 1od}$) resembles (a modified version of) the method proposed in Wang et al.~\cite{Wang:2016:ICDE} that is adopted to fit the visited-POI assignment task.
      Note that CRF(JE$_{\rm 1od}$) remains part of our method since it can be interpreted as a degraded version of JE.
      The difference between JE and CRF(JE$_{\rm 1od}$) is how they calculate the POI-POI features.
      CRF(JE$_{\rm 1od}$) only incorporates the POI-POI features obtained from neighbors in a session.
\end{enumerate}
These six methods all use the same extracted stay-points as candidates of significant locations.

\subsubsection{Experimental settings:}
We used Kang's method as a stay-point extraction algorithm.
As we confirmed in the previous section, our framework assumes exhaustively extracted significant location candidates as input.
Therefore, we used a parameter set of $(\theta_{\rm dist}=100$, $\theta_{\rm time}=180)$.

Basically, we split individual data into two parts, training and test.
For Rank(LI11) and Rank(SH13), the training data were used to build a ranking model.
For JE(full) and CRF(JE$_{\rm 1od}$), we used training data to create features and select weight parameter ${\bf w} \in \{0.1, 1.0\}^W$ with the highest F1-score by a grid-search, where ${\bf w} = ({\bf w}^{(s)}, {\bf w}^{(\bar{s})}, {\bf w}^{(v)}, {\bf w}^{(t)}, {\bf w}^{(y)})$ and $W$ denotes the dimension number of $\bf{w}$.
We used Gurobi Optimizer v.7.5\footnote{\url{http://www.gurobi.com/}} as an ILP solver for JE(full) and CRF(JE$_{\rm 1od}$).
We set $\gamma=0.1$, $\lambda=1/30$, $\alpha_1, \alpha_2 = 0.9$, $\beta=0.01$ for feature calculation\footnote{We selected these parameters based on the results of preliminary experiments on the data obtained in the first round (A01, A02, and A03).}.
Then we used ten-fold cross-validation as the evaluation setting to test the fundamental performance of the visited-POI assignment on several comparative methods.

\subsubsection{Evaluation measure:}
\begin{figure}[t]
\centering
\includegraphics[width=12cm, clip]{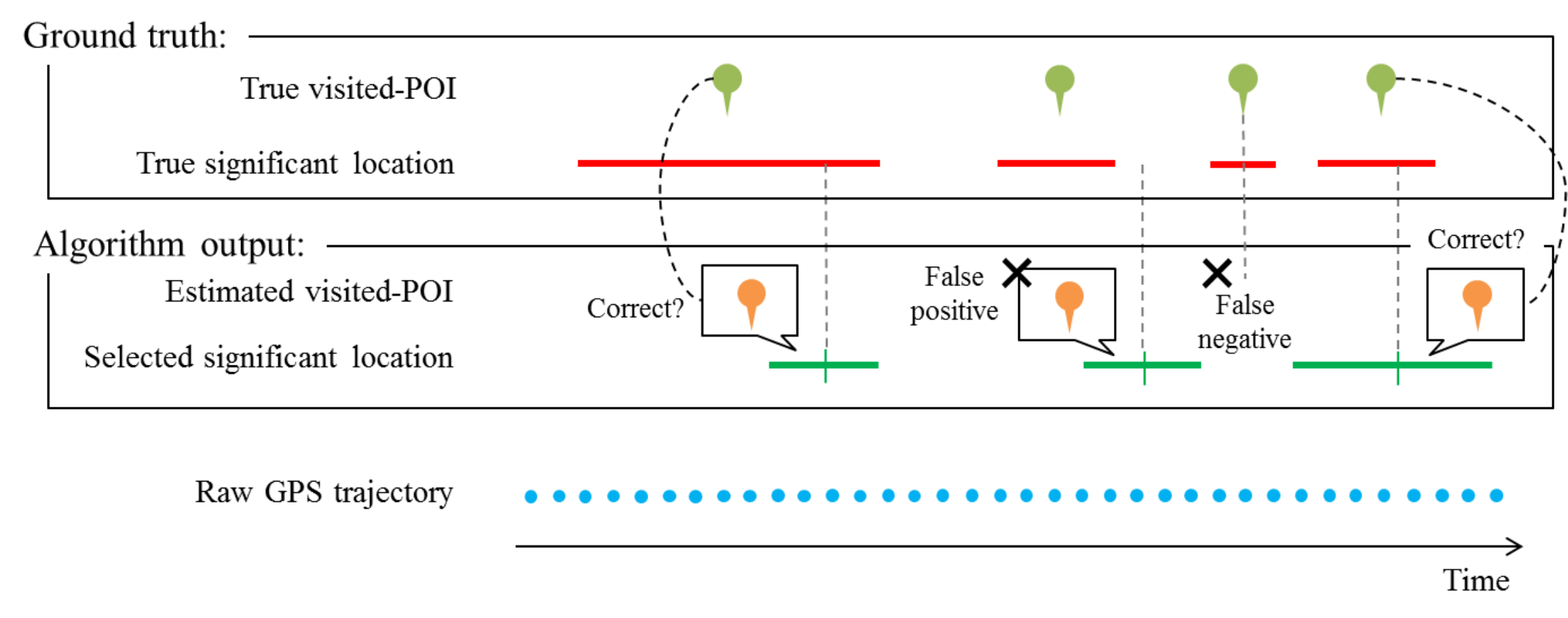}
\caption{Evaluation method for visited-POI assignment task}
\label{fig:evaluation_overview}
\end{figure}

All the experiments in this paper were evaluated based on their F1-score, calculated by the harmonic mean of the precision and recall scores.
Fig.~\ref{fig:evaluation_overview} shows how the F1-score is measured from the true significant locations and their visited-POIs.
Our dataset contains the true significant locations and the corresponding visited-POIs assigned by the users.
In the figure, there are four significant locations and four visited-POIs as ground-truth labels.
A method selects three stay-points and assigns a visited-POI to them in this example.
First, we match a selected stay-point to the true significant location that contains the middle timestamp ($(sp_i.bt + sp_i.et)/2$) of the $i$-th stay-point.
Two of the three selected stay-points match the true significant locations in the example.
If an extracted stay-point matches the true significant location, we can determine whether the predicted visited-POI agrees with the true visited-POI assignment.
The prediction falls under true positive ($TP_{\rm POI}$) if it agrees with the ground-truth label and falls under false positive ($FP_{\rm POI}$) otherwise.
We always judge the selected stay-points that do not match the true significant locations as $FP_{\rm POI}$.
We also assess a true significant location that is not matched by any selected stay-point as a false negative ($FN_{\rm POI}$).

With $TP_{\rm POI}$, $FP_{\rm POI}$ and $FN_{\rm POI}$, we can calculate the precision ($PRE_{\rm POI}$) and recall ($REC_{\rm POI}$) scores in a similar way as in Eq. (\ref{eq:precision_recall}), namely;
\begin{equation}\label{eq:poi_precision_recall}
 PRE_{\rm POI} = \frac{TP_{\rm POI}}{TP_{\rm POI}+FP_{\rm POI}}, \hspace{1em}  REC_{\rm POI} =  \frac{TP_{\rm POI}}{TP_{\rm POI}+FN_{\rm POI}}.
\end{equation}
Then F1-score (F1) is obtained by its harmonic mean;
\begin{equation}\label{eq:poi_f1}
 {\rm F1}_{\rm POI} = \frac{2PRE_{\rm POI} REC_{\rm POI}}{PRE_{\rm POI}+REC_{\rm POI}} =  \frac{2TP_{\rm POI}}{2TP_{\rm POI}+FN_{\rm POI}+FP_{\rm POI}}.
\end{equation}

\begin{figure}[t]
\centering
\includegraphics[width=14cm, clip]{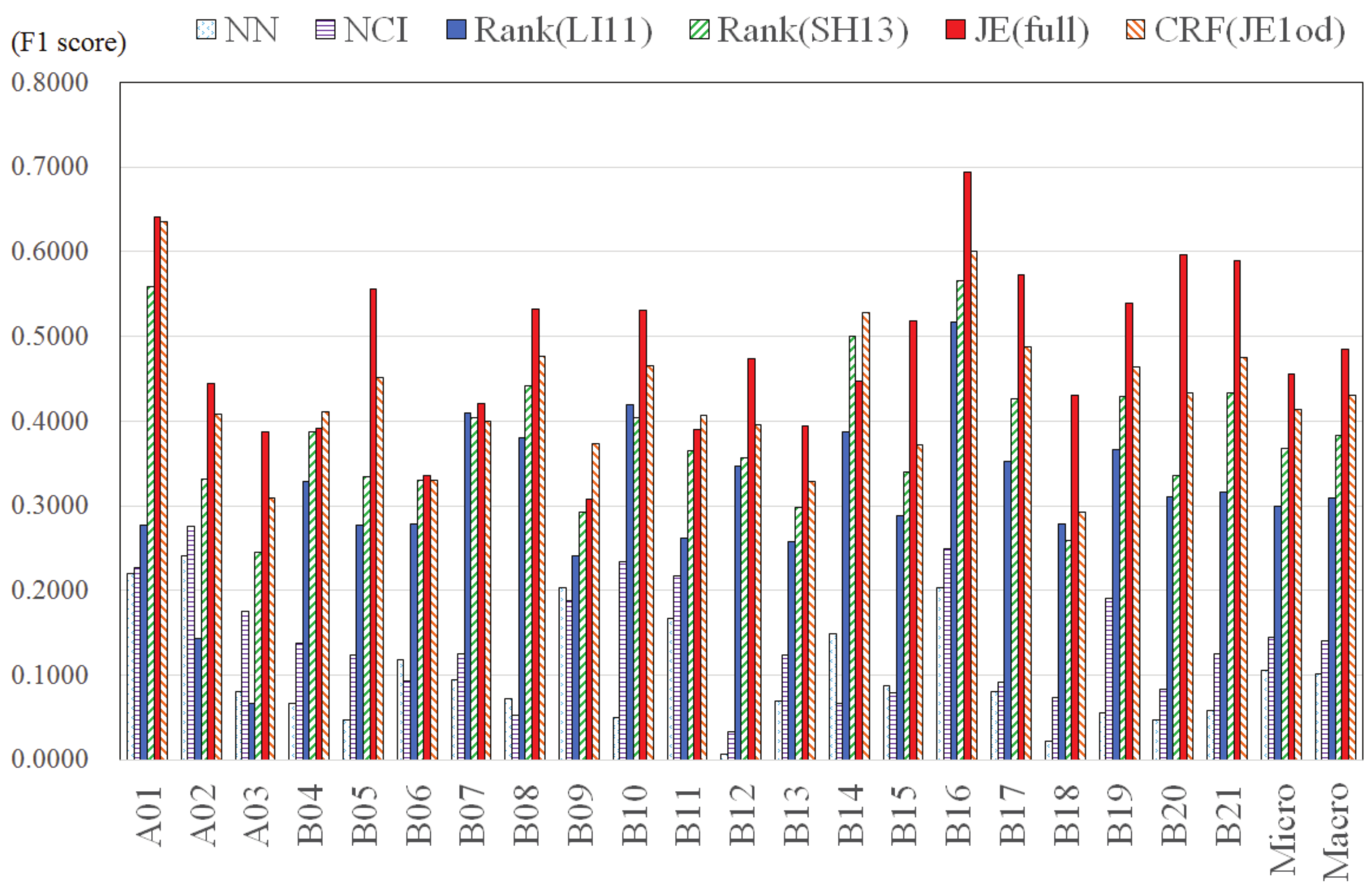}
\caption{F1-scores on ten-fold cross-validation}
\label{fig:result_cv}
\end{figure}

\begin{figure}[t]
\centering
\includegraphics[width=14cm, clip]{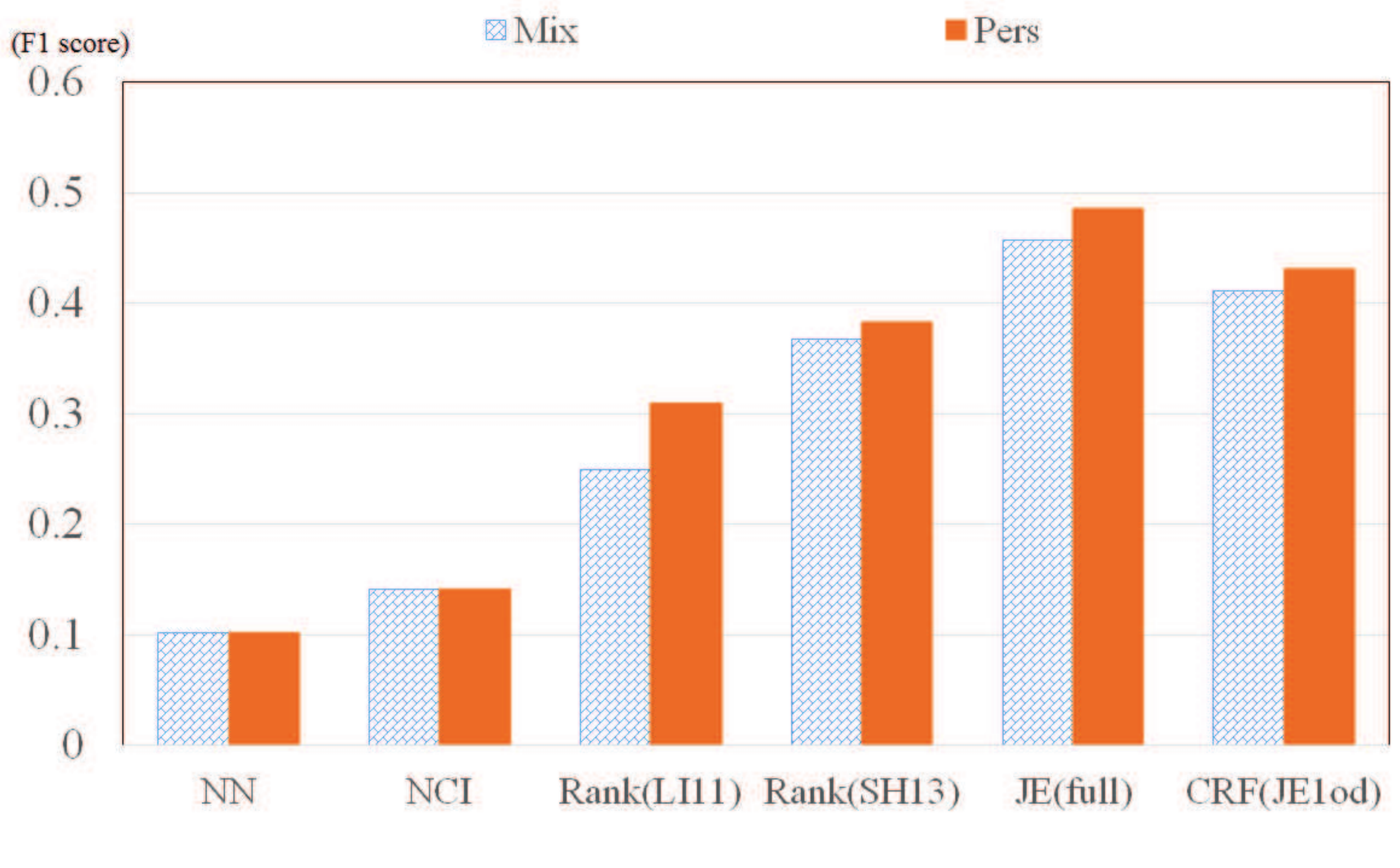}
\caption{F1-scores on mixed and personalized settings for each method}
\label{fig:result_personalized}
\end{figure}

\subsection{Results}
\subsubsection{Performance comparison among comparative methods:}
\label{sec:result1}
Fig.~\ref{fig:result_cv} shows the ten-fold cross-validation results for each individual user.
The x-axis is each user ID, where Micro and Macro respectively represent the micro and macro averages of all the user results.

At first, we observed that JE(full) obtained the best F1 scores on 17 out of 21 users.
We conducted Wilcoxon signed rank test at a confidence level of 0.01 on the Macro averages between all the pairs of JE(full) and other comparative methods, namely (1) JE(full) vs. NN, (2) JE(full) vs. NCI, (3) JE(full) vs. Rank(LI11), (4) JE(full) vs. Rank(SH13), and (5) JE(full) vs. CRF(JE$_{1od}$).
We confirmed that
there were statistical differences on all the five significance tests.
Therefore, we conclude that JE(full) significantly outperformed the other comparative methods.

Next, NN, NCI, Rank(LI11), and Rank(SH13) are point-wise estimation approaches, whereas CRF(JE$_{1od}$) and JE(full) are joint estimation approaches.
In our evaluation, the joint estimation approach is preferable for visited-POI assignments.
This implies that visited-POIs are colereted each other.

Moreover, JE(full) outperformed CRF(JE$_{1od}$) in the same joint estimation approach.
This fact indicates that long-distance dependency information is essential for our task since the difference between JE(full) and CRF(JE$_{1od}$) addresses whether the methods incorporate long-distance transition information.
This explains the need to formulate the task as an ILP problem for easily and efficiently incorporating the information.

\subsubsection{Effectiveness of individual visited-POI assignment setting}
We conducted additional experiments to verify the idea that personal annotation, rather than annotations made by all users, contains essential information to achieve high accuracy in our task.
Fig. \ref{fig:result_personalized} shows the result of each comparative method when we mixed all the user data.
An additional label, Mix, represents the results in the {\it mixed setting};
we trained the models using all the user training data all at once, and evaluated each user's test data by the trained model.
Another additional label, Pers, represents the results in the {\it personalized setting}.
These results are the macro-averages of the F1-scores of the individual visited-POI assignments for each method in Fig.~\ref{fig:result_cv}.

At first, the NN and NCI results are identical since these methods do not perform any training using the training data.
Therefore, the results are always equivalent regardless of the training data.
Then, clearly, the Pers results were consistently better than Mix results.
We also confirmed that results of the Pers setting on JE(full), CRF(JE$_{1od}$) and Rank(LI11) have significant difference on Wilcoxon signed rank test at a confidence level of 0.01 from those of the Mix setting.
Note here that the training data sizes of the personalized setting were approximately 21 times smaller on average than the mixed setting.
However, the personalized setting consistently outperformed the mixed setting.
These are surprising results.
This observation is empirical evidence for the effectiveness of personalized visited-POI assignments.

\subsection{Efficiency analysis}
\begin{figure}[t]
\centering \includegraphics[width=10cm, clip]{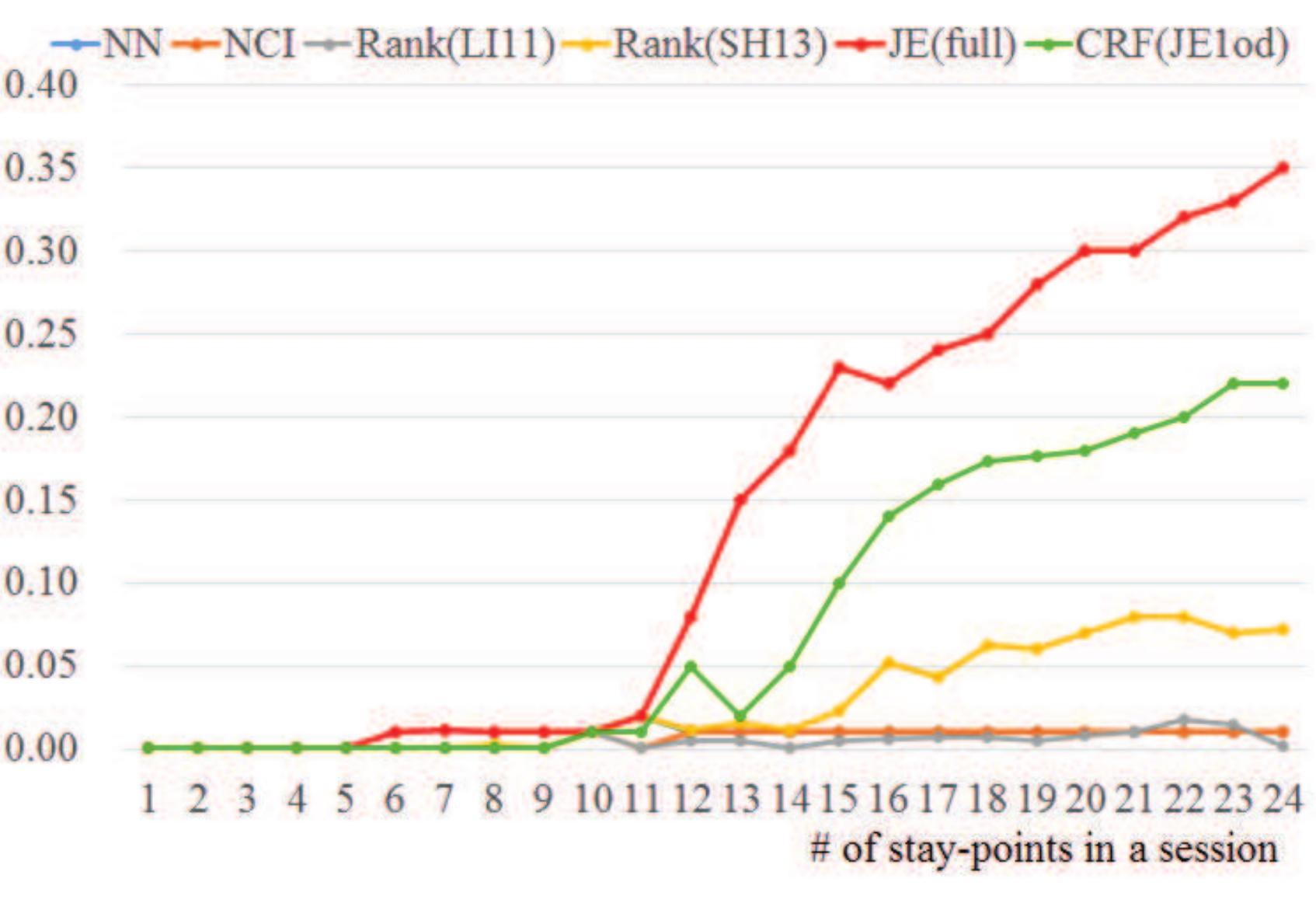}
\caption{Elapsed time over number of stay-points for each comparative method}
\label{fig:ilptime}
\end{figure}

We investigated the actual elapsed time for JE(full) and the other comparative methods since solving ILP problems entails high computational cost.
We measured the average elapsed time for splitting all of the user data into numbers of stay-points.
Fig.~\ref{fig:ilptime} shows the results, where the $x$-axis denotes the number of stay-points in the session and the $y$-axis denotes the actual elapsed time [sec.].
The experiment was conducted on a single machine.

Our joint estimation approach, JE(full), clearly took more calculation time than the other comparative methods.
This observation is expected since JE(full) runs an ILP solver to obtain its result, but the other methods (NN, NCI, and Rank) estimate the results in a point-wise estimation manner.
Therefore, there is a trade-off between calculation cost and performance.

However, here the elapsed time of JE(full) actually averaged less than 0.4 seconds,
suggesting that JE(full) can be run in a practical time.
This is because we only need to run the system once a day in our setting: one second per day for a single user.
Although the current performance of smartphones/tablets is much inferior to that of the machine used in our experiment, the experimental settings remain realistic since smartphones/tablets need a few seconds for processing a single user's private data.
We believe our results confirm the efficiency of our algorithm.

\begin{figure}[t]
\centering
\includegraphics[width=10cm, clip]{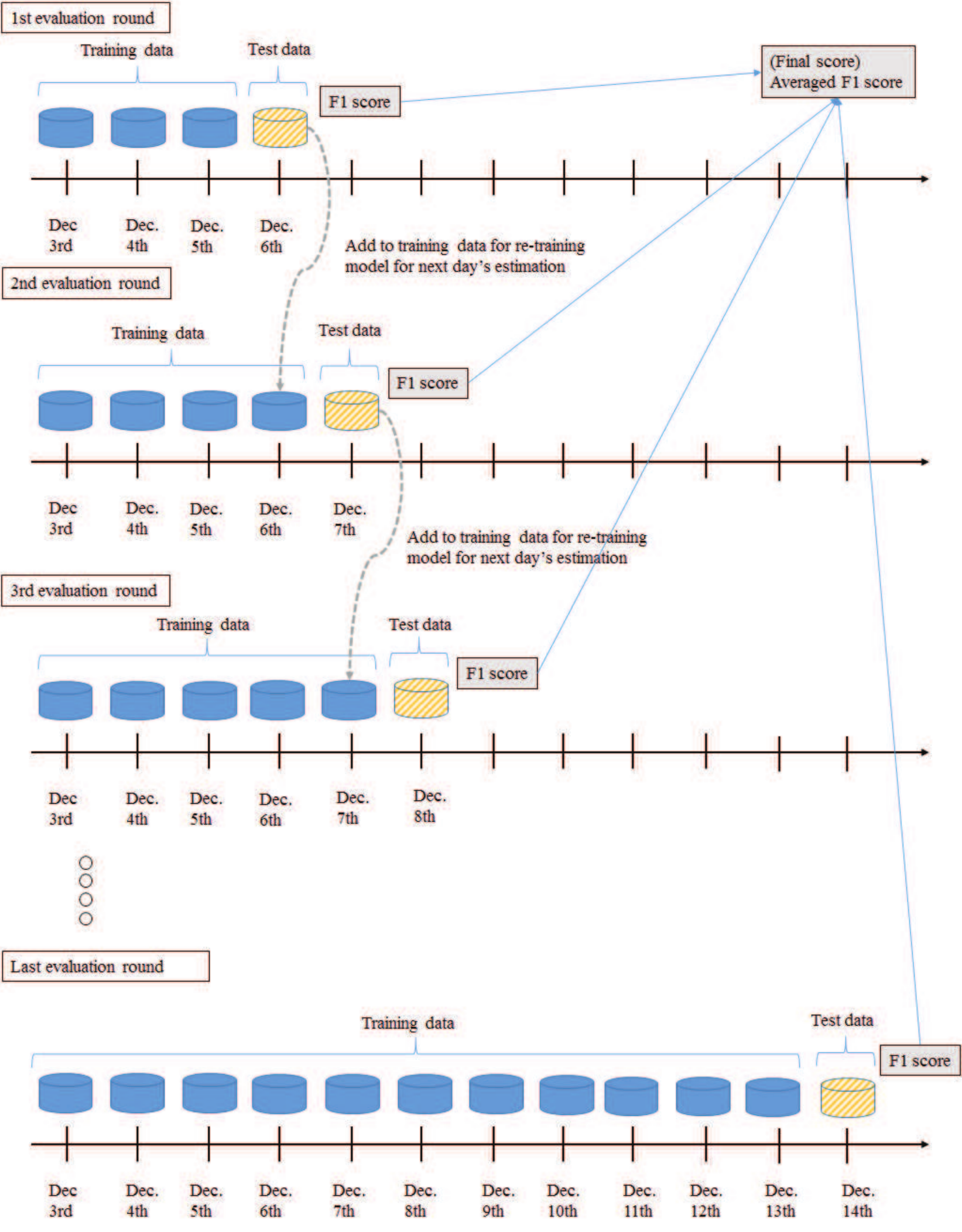}
\caption{Brief sketch of procedure of our sequential evaluation}
\label{fig:evaluation_seq}
\end{figure}
\begin{figure}[t]
\centering
\includegraphics[width=14cm, clip]{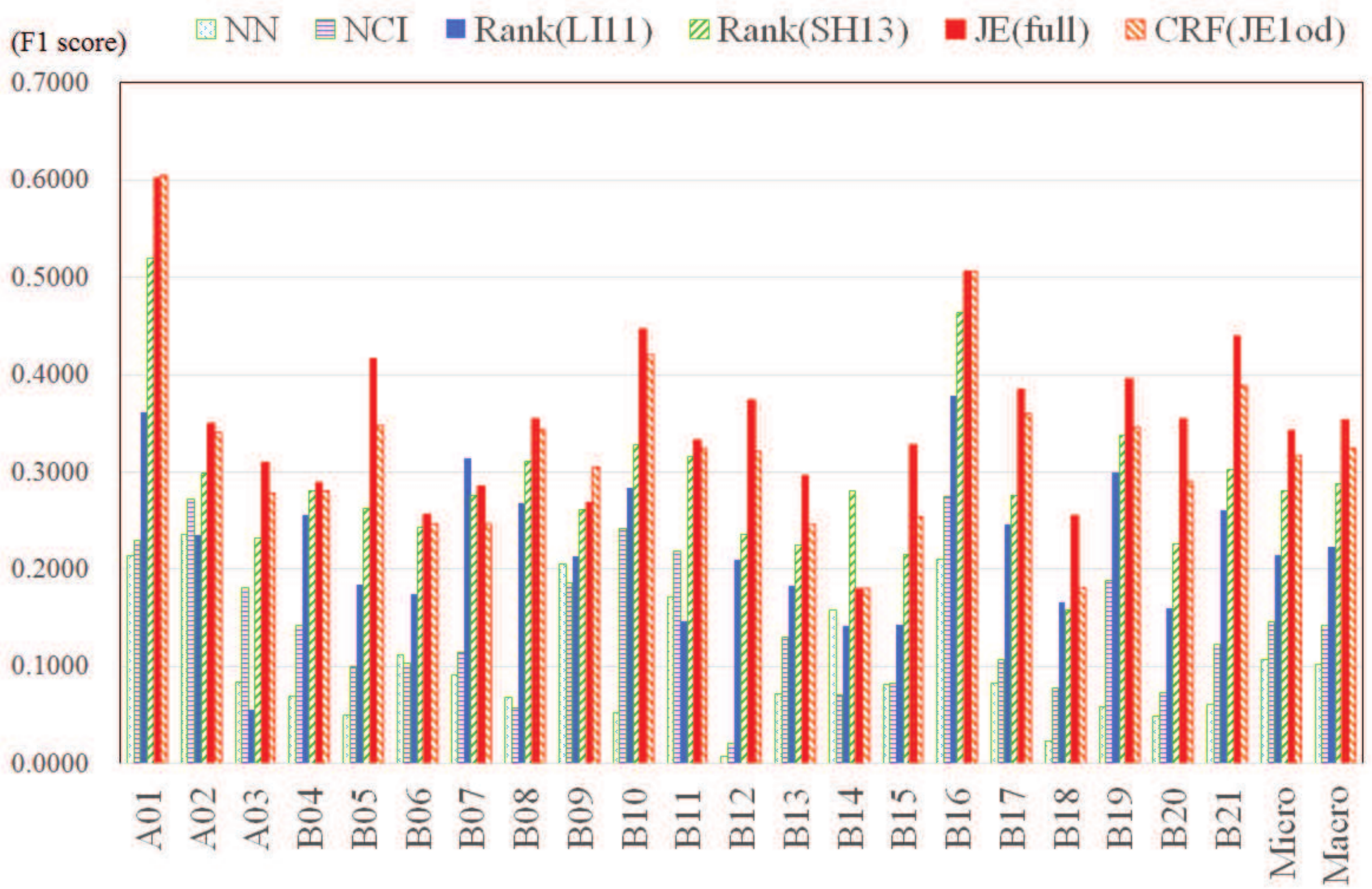}
\caption{Results on sequential evaluation}
\label{fig:result_seq}
\end{figure}

\subsection{Evaluation of sequential visited-POI assignments}
A cross-validation evaluation may not follow the actual use-case of visited-POI assignment tasks.
Therefore, we conducted additional experiments to verify the effectiveness of actual usage.
To resemble the actual usage of the visited-POI assignment, we considered the following procedure as an evaluation setting:
\begin{enumerate}
\item Train a personalized model for each user using the first three sessions (days) as training data.
\item Evaluate the next days' session (the fourth session for first evaluation round, and always evaluate a single session in an evaluation).
\item Re-train (or update) the model by adding the evaluated data used in (2) to the training data.
\item Repeat (2) and (3) alternatively until the last session is evaluated in the procedure (2).
\item Report the average performance of the entire evaluation in the procedure (2).
\end{enumerate}
The above process resembles a daily update system that estimates every session once a day and then updates the model.
We refer to this configuration as a sequential evaluation.
Fig. \ref{fig:evaluation_seq} shows a brief sketch of our sequential evaluation procedure.

Fig.~\ref{fig:result_seq} shows the results on a sequential evaluation setting.
Compared with the results in Fig. \ref{fig:result_cv}, the Micro and Macro average worsened in all the experiments.
This is because the sequential evaluation's task setting is more difficult than the cross-validation since the amount of training data becomes much smaller at the sequential evaluation's beginning.
However, the tendency of the performance gaps among comparative methods resembles those observed in Fig. \ref{fig:result_seq}.
We observed that JE(full) still obtained the best F1 scores on 16 out of 21 users.
We also conducted Wilcoxon signed rank test at a confidence level of 0.01 on the Macro averages between all the pairs of JE(full) and other comparative methods, as described in Section~\ref{sec:result1}, namely (1) JE(full) vs. NN, (2) JE(full) vs. NCI, (3) JE(full) vs. Rank(LI11), (4) JE(full) vs. Rank(SH13), and (5) JE(full) vs. CRF(JE$_{1od}$).
We also confirmed that there were statistical differences on all the five significance tests.
Thus, JE(full) significantly outperformed the other comparative methods on the setting of sequential visited-POI assignments.

\section{Conclusion}
\label{sec:conclusion}
We tackled the problems involved in a personalized visited-POI assignment task that
assigned visited-POIs to their corresponding significant locations in given individual GPS trajectory data with partially annotated user data and a POI database.
We developed a novel visited-POI selection framework based on 0-1 ILP formulation that selects true significant locations and simultaneously assigns visited-POIs while considering different aspects of the selected significant locations and assigned visited-POIs.
Experimental results showed that a conventional stay-point extraction algorithm cannot simultaneously achieve both precision and recall when extracting true significant locations.
Our results also showed that, for performing the visited-POI assignment task, the framework we developed outperforms conventional methods using various cascaded procedures.
Although our method solves an ILP problem that entails high computational cost, we confirmed that it can also output results in a practical time.

\bibliographystyle{plain}

\end{document}